\RequirePackage[switch]{lineno_babar_revtex}
\documentclass[twocolumn,showpacs,aps,pr,nofootinbib]{revtex4}
\catcode`\@=11\relax
\def\@makecol{
 \setbox\@outputbox\vbox{
  \boxmaxdepth\@maxdepth
 \protected@write\@auxout{}{ 
 \string\@LN@col{\@ifnum{\pagegrid@cur=\@ne}{1}{2}}
      }
  \@tempdima\dp\@cclv
  \unvbox\@cclv
  \vskip-\@tempdima
 }
 \xdef\@freelist{\@freelist\@midlist}\global\let\@midlist\@empty
 \@combinefloats
 \@combineinserts\@outputbox\footins
  \set@adj@colht\dimen@
  \count@\vbadness
  \vbadness\@M
  \setbox\@outputbox\vbox to\dimen@{
   \@texttop
   \dimen@\dp\@outputbox
   \unvbox\@outputbox
   \vskip-\dimen@
   \@textbottom
  }
  \vbadness\count@
 \global\maxdepth\@maxdepth
}

\def\balance@two#1#2{
\outputdebug@sw{{\tracingall\scrollmode\showbox#1\showbox#2}}{}
 \setbox\@ne\vbox{
  \@ifvoid#1{}{
   \unvcopy#1\recover@footins
   \@ifvoid#2{}{\marry@baselines}
  }
  \@ifvoid#2{}{
   \unvcopy#2\recover@footins
  }
 }
 \dimen@\ht\@ne\divide\dimen@\tw@
 \dimen@i\dimen@
 \vbadness\@M
 \vfuzz\maxdimen
 \loopwhile{
  \dimen@i=.5\dimen@i
  \outputdebug@sw{\saythe\dimen@\saythe\dimen@i\saythe\dimen@ii}{}
  \setbox\z@\copy\@ne\setbox\tw@\vsplit\z@ to\dimen@
  \setbox\z@ \vbox{
 \protected@write\@auxout{}{ 
   \string\@LN@col{\@ifnum{\pagegrid@cur=\@ne}{1}{2}}
      }
   \unvcopy\z@
   \setbox\z@\vbox{\unvbox\z@ \setbox\z@\lastbox\aftergroup\vskip\aftergroup-\expandafter}\the\dp\z@\relax
  }
  \setbox\tw@\vbox{
   \unvcopy\tw@
   \setbox\z@\vbox{\unvbox\tw@\setbox\z@\lastbox\aftergroup\vskip\aftergroup-\expandafter}\the\dp\z@\relax
  }
  \dimen@ii\ht\tw@\advance\dimen@ii-\ht\z@
  \@ifdim{\dimen@i>.5\p@}{
   \advance\dimen@\@ifdim{\dimen@ii<\z@}{}{-}\dimen@i
   \true@sw
  }{
   \@ifdim{\dimen@ii<\z@}{
    \advance\dimen@\tw@\dimen@i
    \true@sw
   }{
    \false@sw
   }
  }
 }
 \outputdebug@sw{\saythe\dimen@\saythe\dimen@i\saythe\dimen@ii}{}
\@ifdim{\ht\z@=\z@}{
\@ifdim{\ht\tw@=\z@}{
\true@sw
}{
\false@sw
}
}{
\true@sw
}
{
}{
\ltxgrid@info{Unsatifactorily balanced columns: giving up}
\setbox\tw@\box#1
\setbox\z@ \box#2
}
 \setbox\tw@\vbox{\unvbox\tw@\vskip\z@skip}
 \setbox\z@ \vbox{\unvbox\z@ \vskip\z@skip}
 \set@colroom
\dimen@\ht\z@\@ifdim{\dimen@<\ht\tw@}{\dimen@\ht\tw@}{}
\@ifdim{\dimen@>\@colroom}{\dimen@\@colroom}{}
 \outputdebug@sw{\saythe{\ht\z@}\saythe{\ht\tw@}\saythe\@colroom\saythe\dimen@}{}
\setbox#1\vbox to\dimen@{\unvbox\tw@\unskip\raggedcolumn@skip}
\setbox#2\vbox to\dimen@{\unvbox\z@ \unskip\raggedcolumn@skip}
\outputdebug@sw{{\tracingall\scrollmode\showbox#1\showbox#2}}{}
}
\catcode`\@=12\relax

\usepackage{multirow}
\usepackage{graphicx}
\usepackage{dcolumn}
\usepackage{amsmath}
\usepackage{epsfig}
\usepackage{subfigure}

\include{babarsym}

\newcommand{\BABARPubYear}    {}
\newcommand{\BABARPubNumber}  {}

\newcommand{\SLACPubNumber} {}

\newcommand{\comment}[1]{}  

\usepackage{color}
\usepackage{amssymb}

\definecolor{light-gray}{gray}{0.95}
\definecolor{middle-gray}{gray}{0.5}
\definecolor{middle-grey}{gray}{0.5}
\definecolor{gray}{gray}{0.5}
\definecolor{grey}{gray}{0.5}
\definecolor{dark-gray}{gray}{0.25}
\definecolor{dark-grey}{gray}{0.25}
\definecolor{olive}{cmyk}{0.64,0,0.95,0.4}

\def\figurebox#1#2#3{
    \def\arg{#3}
    \ifx\arg\empty
    {\hfill\vbox{\hsize#2\hrule\hbox to #2{\vrule\hfill\vbox to #1{\hsize#2\vfill}\vrule}\hrule}\hfill}
    \else
    {\hfill\epsfbox{#3}\hfill}
    \fi}

\begin{document}
\preprint{\babar-PUB-\BABARPubYear/\BABARPubNumber} 
\preprint{SLAC-PUB-\SLACPubNumber} 
\comment{Changelog:}
\begin{flushleft}
\babar-PUB-12/028\\
SLAC-PUB-15363\\
\end{flushleft}
\title{
{\large \bf
\boldmath
Study of the decay $\Bzb\to\LCp\antiproton\pip\pim$ and its intermediate states
\unboldmath
}
}
%
\author{J.~P.~Lees}
\author{V.~Poireau}
\author{V.~Tisserand}
\affiliation{Laboratoire d'Annecy-le-Vieux de Physique des Particules (LAPP), Universit\'e de Savoie, CNRS/IN2P3,  F-74941 Annecy-Le-Vieux, France}
\author{E.~Grauges}
\affiliation{Universitat de Barcelona, Facultat de Fisica, Departament ECM, E-08028 Barcelona, Spain }
\author{A.~Palano$^{ab}$ }
\affiliation{INFN Sezione di Bari$^{a}$; Dipartimento di Fisica, Universit\`a di Bari$^{b}$, I-70126 Bari, Italy }
\author{G.~Eigen}
\author{B.~Stugu}
\affiliation{University of Bergen, Institute of Physics, N-5007 Bergen, Norway }
\author{D.~N.~Brown}
\author{L.~T.~Kerth}
\author{Yu.~G.~Kolomensky}
\author{G.~Lynch}
\affiliation{Lawrence Berkeley National Laboratory and University of California, Berkeley, California 94720, USA }
% ...
\collaboration{The \babar\ Collaboration}
\noaffiliation

\date{January 29, 2013}

\begin{abstract}
We study the decay $\Bzb\to\LCp\antiproton\pip\pim$, reconstructing the \LCp baryon in the $\proton\Km\pip$ mode,  using a data sample of $467\times 10^{6}$ \BBbar pairs collected with the \babar detector at the \pep2 storage rings at SLAC. We measure branching fractions for decays with intermediate $\SigmaC$ baryons to be $\BR[\Bzb\to\SigmaC(2455)^{++}\antiproton\pim]=( 21.3 \pm 1.0\comment{_{\mathrm{stat}}} \pm 1.0\comment{_{\mathrm{sys}}}  \pm 5.5\comment{_{\LCp}})  \times 10^{-5}$, $\BR[\Bzb\to\SigmaC(2520)^{++}\antiproton\pim]=(11.5\pm 1.0\comment{_{\mathrm{stat}}} \pm 0.5\comment{_{\mathrm{sys}}}  \pm 3.0\comment{_{\LCp}} )\times 10^{-5}$, $\BR[\Bzb\to\SigmaC(2455)^{0}\antiproton\pip]=(9.1 \pm 0.7\comment{_{\mathrm{stat}}}  \pm 0.4\comment{_{\mathrm{sys}}}   \pm 2.4\comment{_{\LCp}})\times10^{-5}$, and $\BR[\Bzb\to\SigmaC(2520)^{0}\antiproton\pip]= \left(2.2 \pm 0.7 \pm 0.1\pm 0.6\right) \times 10^{-5}$, where the uncertainties are statistical, systematic, and due to the uncertainty on the $\LCp\to\proton\Km\pip$ branching fraction, respectively. For decays without $\SigmaC(2455)$ or $\SigmaC(2520)$ resonances, we measure  $\BR[\Bzb\to\LCp\antiproton\pip\pim]_{\mathrm{non-\SigmaC}}=(79  \pm 4\comment{_{\mathrm{stat}}} \pm 4\comment{_{\mathrm{sys}}}  \pm 20\comment{_{\LCp}})\times10^{-5}$. The total branching fraction is determined to be $\BR[\Bzb\to\LCp\antiproton\pip\pim]_{\mathrm{total}}=(123  \pm 5\comment{_{\mathrm{stat}}} \pm 7\comment{_{\mathrm{sys}}}  \pm 32\comment{_{\LCp}})\times10^{-5}$.
We examine multibody mass combinations in the resonant three-particle $\SigmaC\antiproton\pi$ final states and in the four-particle $\LCp\antiproton\pip\pim$ final state, and observe different characteristics for the $\antiproton\pi$ combination in neutral versus doubly-charged $\SigmaC$ decays.
\end{abstract}
\maketitle
\section{Introduction}
Decays of \B mesons into final states with baryons account for $\left(6.8\pm0.6\right)\%$ \cite{PDG2010} of all \B meson decays. Notwithstanding their significant production rate, the baryon production mechanism in \B meson decays is poorly understood. 
Theoretical models of \B meson baryonic decays are currently limited to rough estimates of the  branching fractions and basic interpretations of the decay  mechanisms \cite{Savage:1989jx,ref:Soni,Cheng:2001ub,Cheng:2002sa,Suzuki:2006nn}. Additional experimental information may help to clarify the underlying dynamics.\\\indent
In this paper, we present a measurement of the \B-meson baryonic decay\footnote{The use of charge conjugate decays is implied throughout this paper.} $\Bzb\to\LCp\antiproton\pip\pim$. The \LCp baryon is observed through its decays to the $\proton\Km\pip$ final state. The study is performed using a sample of \epem annihilation data collected at the mass of the \Y4S resonance with the \babar detector at the SLAC National Accelerator Laboratory.  We include a study of the production of this final state through intermediate \SigmaCplpl and \SigmaCz resonances. The \sPlot technique \cite{Pivk:2004ty} is used to examine multibody mass combinations within the $\SigmaC\antiproton\pi$ final states. We account for background from sources such as $\B\to D\proton\antiproton \, \left({{n}}\pi\right)$ and $\Bm\to\SigmaCpl\antiproton\pim$, which were not considered in previous studies \cite{Lit:Dytman:2002yd,BELLE_SigmaC}. In addition, we extract the four-body non-resonant branching fraction and examine two- and three-body mass combinations within the four-body $\LCp\antiproton\pip\pim$ final state. The $\Bzb\to\LCp\antiproton\pip\pim$ decay has previously been studied by the CLEO \cite{Lit:Dytman:2002yd} and Belle \cite{BELLE_SigmaC} Collaborations using data samples of 9.17\invfb of 357\invfb, respectively. The present work represents the first study of this decay mode from \babar. \\\indent
Section \ref{txt:BabarDescription_1} provides a brief description of the \babar detector and data sample. The basic event selection procedure is described in Sec. \ref{txt:EventSelection_1}. Section \ref{txt:SigmaCplplz_1} presents the method used to extract results for channels that proceed via intermediate \SigmaC baryons.  The corresponding results for channels that do not proceed via \SigmaC baryons are presented in Sec. \ref{txt:nonSigmaC_1}. Section \ref{txt:Efficiency_1} presents the method used to determine signal reconstruction efficiencies, Sec. \ref{txt:BranchingRatios_1}  the branching fraction results, Sec. \ref{txt:systematics_1} the evaluation of systematic uncertainties, and Sec. \ref{txt:Results_1} the final results.  A summary is given in Sec. \ref{txt:Summary_1}.

\section{\babar Detector and Data Sample}\label{txt:BabarDescription_1}
The data sample used in this analysis was collected with the \babar detector at the \pep2 asymmetric-energy \epem storage ring at SLAC. \pep2 operates with a 9\gev \en and a 3.1\gev \ep beam resulting in a center-of-mass energy equal to the \FourS mass of 10.58\gevcc. The collected data sample contains $467\times 10^{6}$ \BBbar pairs, which corresponds to an integrated luminosity of $426\invfb$.\\\indent
The \babar detector \cite{Aubert:2001tu} measures charged-particle tracks with a five-layer double-sided silicon vertex tracker (SVT) surrounded by a 40-layer drift chamber (DCH). Charged particles are identified using specific ionization energy measurements in the SVT and DCH, as well as Cherenkov radiation measurements in an internally reflecting ring imaging Cherenkov detector (DIRC). These detectors are located within the $1.5\mathrm{\,T}$ magnetic field of a superconducting solenoid.\\\indent
Using information from the SVT, the DCH, and the DIRC for a particular track, the probability for a given particle hypothesis is calculated from likelihood ratios. The identification efficiency for a proton is larger than 90\% with the probability of misidentifying a kaon or pion as a proton between 3\% and 15\% depending on the momentum. For a kaon, the identification efficiency is 90\% with the probability of misidentifying a pion or proton as a kaon between 5\% and 10\%. The identification efficiency for a pion is larger than 95\% with the probability of misidentifying a kaon or proton as a pion between 5\% and 30\%.\\\indent
Monte Carlo (MC) simulated events are produced with an $\epem\to\BBbar$ event simulation based on the {\tt EvtGen} program \cite{Lange:2001uf} and an $\epem\to\uubar, \ddbar, \ssbar,\ccbar$ event simulation based on the {\tt JETSET} program \cite{Sjostrand:1993yb}. Generated events are processed in a {\tt GEANT4} \cite{Agostinelli:2002hh} simulation of the \babar detector. MC-generated events are studied for generic background contributions as well as for specific signal and background modes. Baryonic \B meson decays are generated assuming that their daughters are distributed uniformly in phase space.

\section{Event Selection}\label{txt:EventSelection_1}
\begin{figure}[h]
        \centering
          \includegraphics[width=0.5\textwidth]{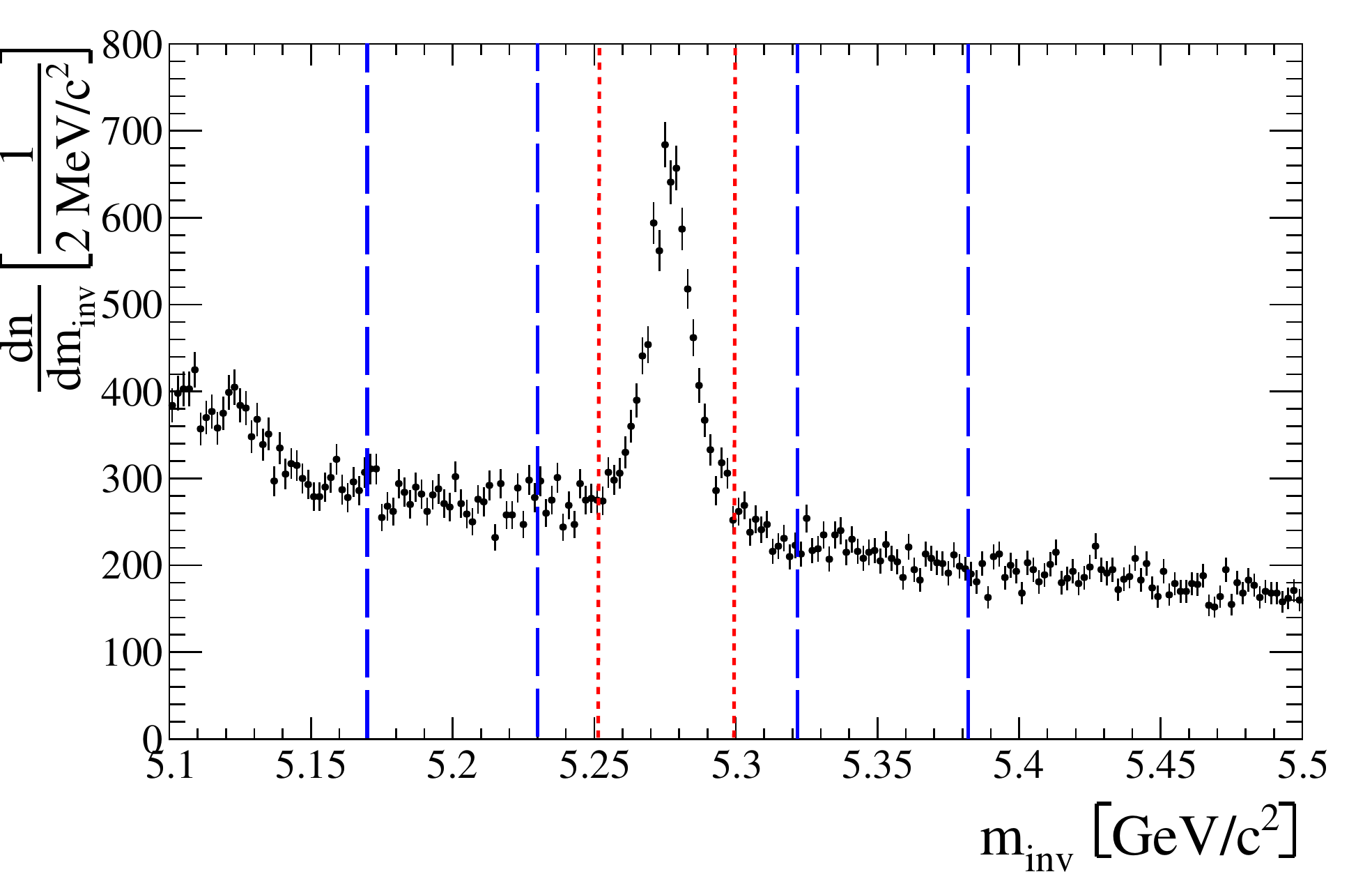}
          \caption{Distribution of the invariant mass $\minvn\left(\LCp\antiproton\pip\pim\right)$ for events with \mes in the region $\left[5.272,5.285\right]\gevcc$. The red dotted lines indicate the signal region and the blue dashed lines the sideband regions. Higher multiplicity modes, such as $\B\to\LCp\antiproton\pip\pim\pi$, appear for $\minvn<5.14\gevcc$.}
          \label{fig:InvMass1}
 
\end{figure}

The signal mode is reconstructed in the decay chain $\Bzb\to\LCp\antiproton\pip\pim$ with $\LCp\to\proton\Km\pip$. All final state particles are required to have well defined tracks in the SVT and DCH. Kaons and protons, as well as pions from the \LCp decay, are required to pass likelihood selectors based on information from the SVT, DCH, and DIRC. For pion candidates from the \Bzb decay, a well reconstructed track is required.\\\indent
To form a $\LCp$  candidate, the \proton, \Km, and \pip candidates are fitted to a common vertex and a \chiq probability greater than $0.1\%$ is required for the vertex fit. To form a \Bzb  candidate, the \LCp  candidate is constrained to its nominal mass value and combined with an antiproton and two pions with opposite charge. The mass constraint value differs between events from data and MC. For the MC events a nominal \LCp  mass of ${m}_{\LCp}^{\mathrm{MC}}=2284.9\mevcc$ is chosen; this corresponds to the mass value used in the MC generation and to the value from fits to reconstructed MC events. 
For data, \chiq fits are performed on the $m\left(\proton\Km\pip\right)$ invariant mass distribution to find the nominal \LCp mass. The fits are performed for each of the six distinct \babar run periods. The results are found to vary between $m_{\LCp}^{\mathrm{data}}=\left(2285.55\pm0.18\right)\mevcc$ and $m_{\LCp}^{\mathrm{data}}=\left(2285.62\pm0.22\right)\mevcc$, where the uncertainties are statistical. All invariant mass values are found to be consistent. The average result ${m}_{\LCp}^{\mathrm{data}}=2285.6\mevcc$ is used as the nominal value for the mass constraint in data.\\\indent
Only candidates within a 25\mevcc mass window centered on the nominal \LCp mass ${m}_{\LCp}^{\mathrm{data}}$ (or ${m}_{\LCp}^{\mathrm{MC}}$ for simulated events) are retained. The entire decay chain is refitted requiring that the direct \Bzb  daughters originate from a common vertex and that the \chiq  probability for the \Bzb  vertex fit exceeds $0.1\%$.\\\indent
The decays  $\B\to D\proton\antiproton \, \left({{n}}\pi\right)$ with ${{n}}=1,2$, which are described in more detail in section \ref{txt:backgrounds_1}, can contribute a signal-like background through rearrangement of the final-state particles and are denoted ``peaking background'' in the following. To suppress these events, symmetric vetoes of $\pm20\mevcc$ around the nominal $\Dz$ and $\Dp$ mass values \cite{PDG2010} are applied in the distributions of the invariant masses $m\left(\left[\Km\pip\right]_{\LCp}\left[\pim\pip\right]_{\Bzb}\right)$, $m\left(\left[\Km\pip\right]_{\LCp}\left[\pip\right]_{\Bzb}\right)$, and $m\left(\left[\Km\pip\right]_{\LCp}\right)$, where subscripts denote the mother candidate of the particles.\\\indent
To separate \Bzb signal events from combinatorial background, two variables are used. The \Bzb invariant mass is defined as $\minvn=\sqrt{E_{\Bzb}^{2}-{\mathbf{p}}_{\Bzb}^{2}}$  with the four-momentum vector of the \Bzb  candidate $\left(E_{\Bzb},{\mathbf{p}}_{\Bzb}\right)$ measured in the laboratory frame. The energy-substituted mass is defined in the laboratory frame as $\mes=\sqrt{{\left(s/2 + {\mathbf{p}}_{i}\cdot{\mathbf{p}}_{\Bzb}\right)^{2}}/{E_{i}^{2}}-{\mathbf{p}}_{\Bzb}^{2}}$ with $\sqrt{s}$ the center-of-mass energy and $\left(E_{i},{\mathbf{p}}_{i}\right)$ the four-momentum vector of the initial \epem  system measured in the laboratory frame. For both variables, genuine \Bzb  decays are centered at the \Bzb  meson mass. In MC, these variables exhibit a negligible correlation for genuine \Bzb  mesons.\\\indent
To suppress combinatorial background, \Bzb  candidates are required to satisfy $\mes\in\left[5.272,5.285\right]\gevcc$. Figure \ref{fig:InvMass1} shows the \minv distribution after applying all of the above selection criteria. The dashed lines show sideband regions $m_{\mathrm{inv}}\in\left[5.170,5.230\right]$ and $m_{\mathrm{inv}}\in\left[5.322,5.382\right]$, used to study background characteristics; both sideband regions are combined into a single sideband region.\\\indent
The analysis is separated into two parts: I) the measurement of the four signal decays via intermediate $\SigmaC(2455,2520)$ resonances\ie $\Bzb\to\SigmaC(2455)^{++}\antiproton\pim$, $\Bzb\to\SigmaC(2520)^{++}\antiproton\pim$, $\Bzb\to\SigmaC(2455)^{0}\antiproton\pip$, and $\Bzb\to\SigmaC(2520)^{0}\antiproton\pip$, and II) the measurement of all other decays into the four-body final state $\LCp\antiproton\pip\pim$, which are denoted as {\it{}non-\SigmaC signal events} in the following.

\begin{figure*}[htpb]
  \centering
     \subfigure{
          \includegraphics[width=0.485\textwidth]{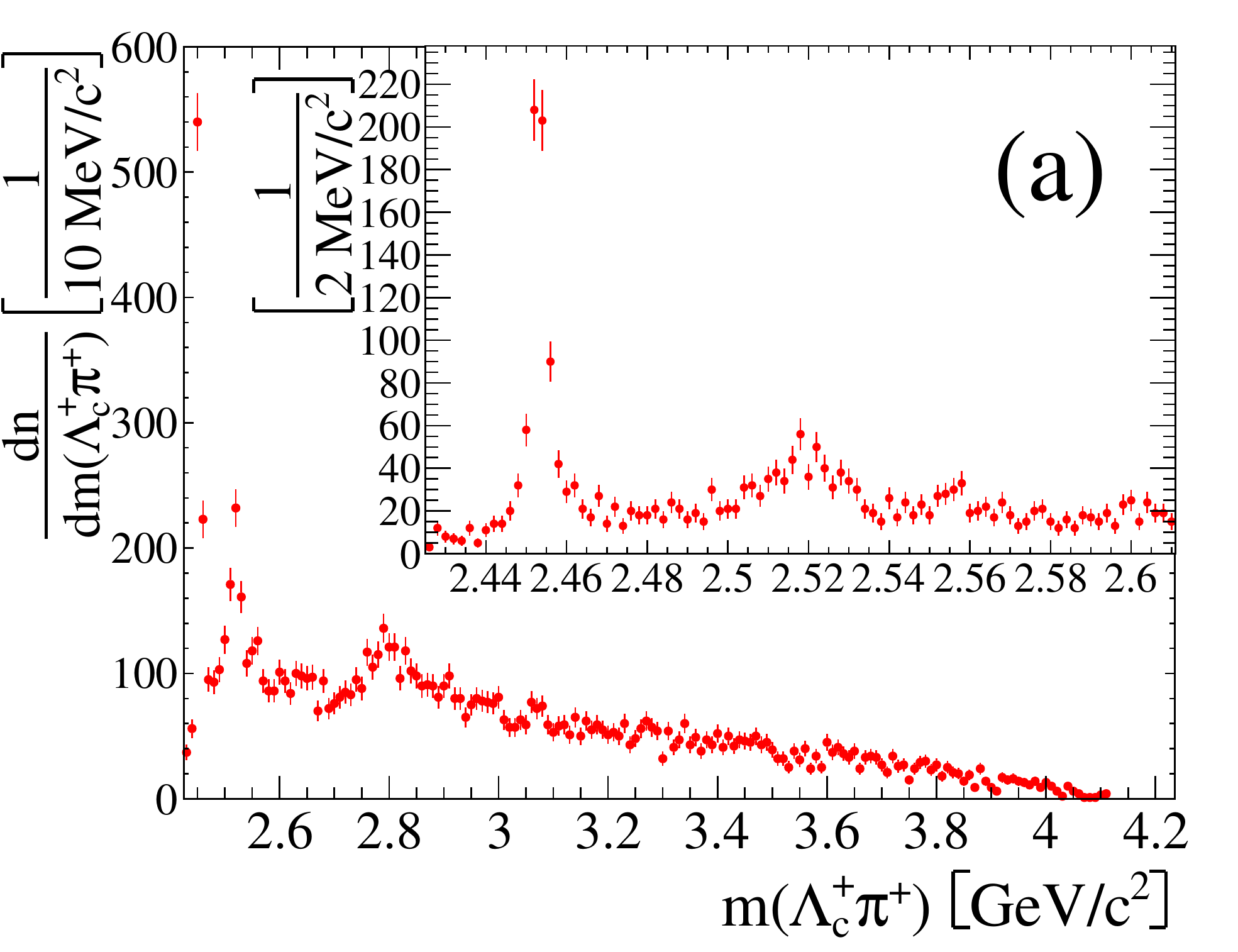}
          \label{fig:SigmaC0plpl1a}
        }
    \subfigure{
          \includegraphics[width=0.485\textwidth]{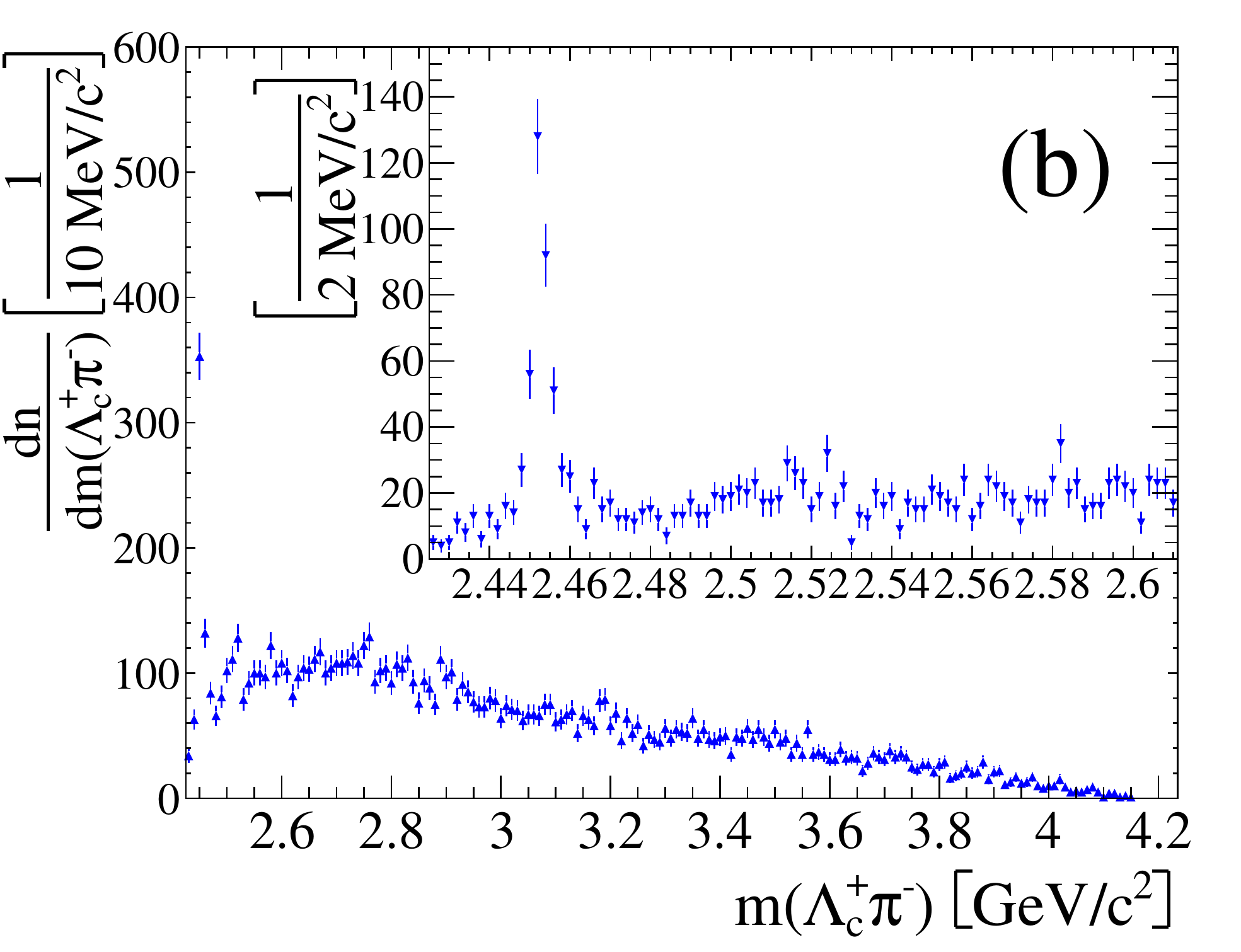}
          \label{fig:SigmaC0plpl1b}
        }
  \caption{Event distributions in $m\left(\LCp\pip\right)$ (a) and  $m\left(\LCp\pim\right)$ (b)  for events in the signal region of Fig. \ref{fig:InvMass1}. The inserts show the low invariant mass regions.}
  \label{fig:SigmaC0plpl1}
\end{figure*}

\begin{boldmath}
\section{$\Bzb\to\SigmaCplplz\antiproton\pi^{\mp}$ Analysis}\label{txt:SigmaCplplz_1}
\end{boldmath} 
Decays via resonant intermediate states with $\SigmaC$ resonances are studied in the two-dimensional planes spanned by \minv and the invariant \SigmaC  candidate invariant mass $\mLCpip$ for decays with ${\SigmaC(2455,2520)^{++}}$ and $\mLCpim$ for decays with $\SigmaC(2455,2520)^{0}$. In the following the like-sign $\LC\pi$ invariant mass is denoted as \mpp and the opposite-sign invariant mass as \mpm. If both invariant masses are referred to, we use the notation \mpppm. For intermediate $\SigmaC(2455,2520)^{++,0}$ states, $\BR\left[\SigmaC\to\LCp\pi\right]\approx100\%$ is assumed \cite{PDG2010}.\\\indent
We perform fits in both planes $\minvn\,{:}\,\mpp$ and  $\minvn\,{:}\,\mpm$ to extract the signal yields for the decays via the \SigmaC resonances. Background contributions are vetoed when feasible. We distinguish between different signal and remaining background contributions by using separate probability density functions (PDF) for each signal and background component. We use analytical PDFs as well as discrete histogram PDFs. The PDFs  are validated using data from the sideband regions and from MC samples. The different, combined PDFs are fitted to the $\minvn\,{:}\,\mpp$ and $\minvn\,{:}\,\mpm$ planes and the resulting covariance matrices of the fits are used to calculate \sPlot \cite{Pivk:2004ty} distributions of signal events.\\\indent
Figures \ref{fig:SigmaC0plpl1}(a)  and \ref{fig:SigmaC0plpl1}(b) show the \mpp and \mpm distributions, respectively, after applying the selection criteria as described in Sec. \ref{txt:EventSelection_1}. Signal contributions from the $\SigmaC(2455)^{++}$, $\SigmaC(2520)^{++}$, and $\SigmaC(2455)^{0}$ resonances are observed and a contribution from events with a $\SigmaC(2800)^{++}$ resonance is visible. The doubly-charged \SigmaCplpl resonances are seen to contribute larger numbers of events than the neutral \SigmaCz resonances. The resonant structures sit on top of combinatorial background and peaking background events as well as non-\SigmaC signal events. The latter are distributed in $m\left(\LCp\pipm\right)$ like combinatorial background events.

\subsection{Background sources}\label{txt:backgrounds_1}
The main source for combinatorial background events is other \B decays, while $20\%$ originate from $\epem\to\ccbar$ events. Combinatorial events do not exhibit peaking structures in the distributions of the signal variables under study. In contrast, other sources of background do exhibit peaking structures, and are treated separately. 
\subsubsection{$\B\to D \proton\antiproton\; \left({n}\pi\right)$}\label{txt:backgrounds_D}
\begin{figure*}[htpb]
        \centering
          \includegraphics[width=0.95\textwidth]{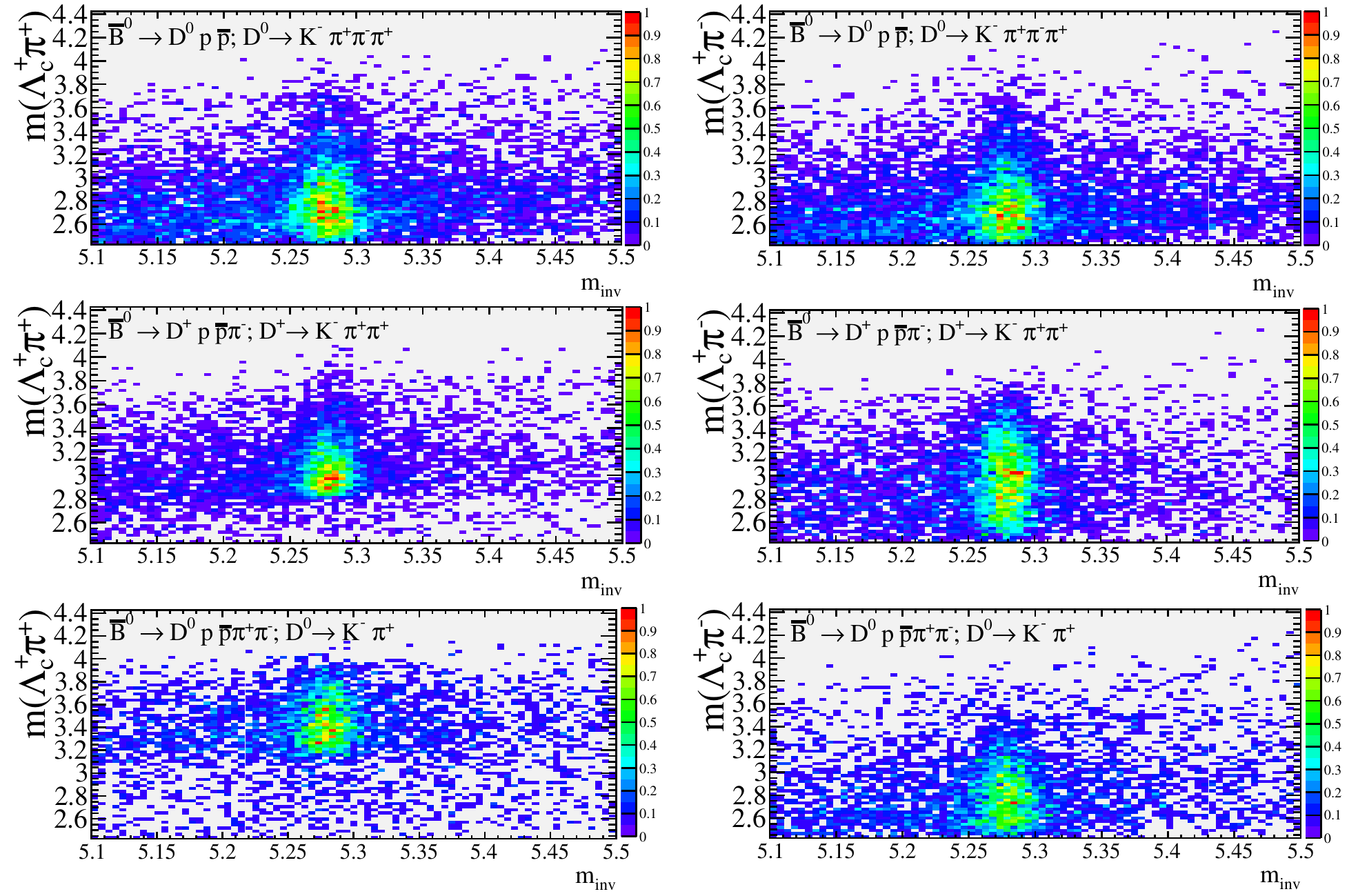}
          \caption{Simulated events with  $\Bzb \to D \proton\antiproton\;\left(n\pi\right)$ decays misidentified as signal decays in the $\minvn\,{:}\,\mpp$ plane (left column) and $\minvn\,{:}\,\mpm$ plane (right column). The MC-generated events are reconstructed as $\Bzb\to\LCp\antiproton\pip\pim$. The color scale indicates the relative contents of a bin compared to the maximally occupied bin (color online).}
          \label{fig:FakeD0Dpl_1}
 \end{figure*}

Decays of the type $\B\to D \proton\antiproton \left({n}\pi\right)$ with  $ D \to\Km  \left({m}\pi\right)$,  where $n+m=3$, can have the same final state particles as signal decays. Rearrangement of the final state particles can yield a fake \LCp  candidate, while the \Bzb  candidate is essentially a genuine \Bzb suppressed only by the \LCp  selection. Because these events represent fully reconstructed genuine \B-meson decays, they are distributed like signal events in the  \minv and \mes variables.  Table \ref{tab:PseudoMinvBG_1} shows the relevant decay modes and their misreconstruction rate as signal. Furthermore, these events can also be misreconstructed as higher \SigmaC  resonances in the $\LCp\pi$ invariant masses. Figure \ref{fig:FakeD0Dpl_1} shows the 
distributions of the MC-simulated background modes in the  $\minvn\,{:}\,\mpp$ and $\minvn\,{:}\,\mpm$ planes. Additionally, $\Bzb \to \Dp \proton\antiproton\pim$ events with $\Dp\to\Km\pip\pip$ have a minimum invariant mass in \mpp of $m\left(\Dp\proton\right)\approx 2.808\gevcc$ and can introduce background in the study of events with intermediate $\SigmaC(2800)^{++}$ resonances.\\\indent
From the misreconstruction efficiency determined from signal MC events and scaled with the measured branching fractions \cite{TaeMin_arXiv}, $167 \pm 20$ background events are expected to contribute as signal. To suppress these events, veto regions are set to 20\mevcc around the nominal \Dz and \Dp masses \cite{PDG2010} in $m\left(\Km\pip\right)$, $m\left(\Km\pip\pip\right)$, and $m\left(\Km\pip\pip\pim\right)$, with the resulting suppression rates given in table \ref{tab:PseudoMinvBG_1}. A systematic uncertainty is assigned to account for the remaining background events. No distortions are found in other variables due to the vetoes. Note that $\Bzb \to \Dz \proton\antiproton\pip\pim$ events with $\Dz\to\Km\pip$ rearranged to $\Bzb\to\bigl[\pim\pip\antiproton\textcolor{black}{\biggl[}\proton\bigr]_{\Bzb}\bigl[\Km\pip\bigr]_{\Dz}\textcolor{black}{\biggr]_{{{\mathrm{fake}\,}\LCp}}}$ do not contribute peaking background because the selection requirement on $m\left(\proton\Km\pip\right)_{\LCp}$ effectively vetoes these events.
\subsubsection{$\Bzb\to\left(\c\cbar\right)\Kstarzb\pip\pim$}\label{txt:backgrounds_Charmonia}
Decays via charmonia, such as $\Bzb\to\left(\c\cbar\right)\Kstarzb\pip\pim$ with $\left(\c\cbar\right)\to\proton\antiproton$ and $\Kstarzb\to\Km\pip$, or  $\Bzb\to\left(\c\cbar\right)\Kstarzb$ with $\left(\c\cbar\right)\to\proton\antiproton\pip\pim$, can also produce the same final state particles as signal events. We observe no indication of such contributions in data in the relevant combinations of \Bzb daughters or in signal MC events when scaling the misreconstruction efficiencies with the measured branching fractions \cite{PDG2010}. We neglect these events, but assign a corresponding systematic uncertainty (see Sec. \ref{txt:systematics_1}).
\begin{table*}[htpb]
  \centering
\caption{Efficiencies for reconstructing $\Bzb \to D \proton\antiproton\,\left({ n} \pi\right)$ events as signal decays by rearranging the final-state particles in signal-like combinations. In the fake signal reconstruction, the subscript particles denote the actual mother.  The quantity ${n_{\mathrm{expected}}}$ gives the number of fake signal events without the $D$-meson veto (see text), $\mathrm{\varepsilon_{\mathrm{Cut}}}$ gives the efficiencies of the vetoes, and ${n_{\mathrm{remaining}}}$ gives the expected number of remaining fake events in the signal regions after applying the vetoes. The $\Bzb \to D \proton\antiproton\; \left( {n} \pi\right)$ branching fractions are taken from Ref. \cite{TaeMin_arXiv} and the \Dz/\Dp branching fractions from Ref. \cite{PDG2010}.\vrule width0pt depth12pt}
\begin{tabular}{c c c c c c}\hline\hline
{ Decay mode} & { Fake signal} & { $\mathrm{\varepsilon}_{\Bzb\to\LCp\antiproton\pip\pim}$} & { ${n_{\mathrm{expected}}}$}  & { $\mathrm{\varepsilon_{\mathrm{Cut}}}$} & { ${n_{\mathrm{remaining}}}$}\\\hline
\begin{minipage}[H]{30mm}\vspace{5 pt}$\Bzb \to \Dz \proton\antiproton$\\$\Dz\to\Km\pip\pim\pip$\vspace{5 pt}\end{minipage} & $\Bzb_{{\mathrm {fake}}} \to \bigl[\antiproton\textcolor{black}{\biggl[}\proton\bigr]_{\Bzb}\bigl[\Km\pip\textcolor{black}{\biggr]_{{{\mathrm{fake}\,}\LCp}}}\pim\pip\bigr]_{\Dz}$ & $(6.79 \pm 0.19)\times 10^{-3}$  & $ 26.0$ & 99.3\% & 0.3 \\[10 pt]
\begin{minipage}[H]{30mm}\vspace{5 pt}$\Bzb \to \Dp \proton\antiproton\pim$\\$\Dp\to\Km\pip\pip$\vspace{5 pt}\end{minipage} &  $ \Bzb_{{\mathrm {fake}}} \to \bigl[\pim\antiproton\textcolor{black}{\biggl[}\proton\bigr]_{\Bzb}\bigl[\Km\pip\textcolor{black}{\biggr]_{{{\mathrm{fake}\,}\LCp}}}\pip\bigr]_{\Dp}$ & $(7.28 \pm 0.17)\times 10^{-3}$  & $ 103.0$ & 98.8\% & 1.0 \\[10 pt]
\begin{minipage}[H]{30mm}\vspace{5 pt}$\Bzb \to \Dz \proton\antiproton\pip\pim$\\$\Dz\to\Km\pip$\vspace{5 pt}\end{minipage} & $ \Bzb_{{\mathrm {fake}}}\to\bigl[\pim\antiproton\textcolor{black}{\biggl[}\proton\pip\bigr]_{\Bzb}\bigl[\Km\textcolor{black}{\biggr]_{{{\mathrm{fake}\,}\LCp}}}\pip\bigr]_{\Dz}$ & $(4.19 \pm 0.15)\times 10^{-3}$  & $ 22.5$ & 96.9\% & 0.2\\[10 pt]
\begin{minipage}[H]{30mm}\vspace{5 pt}$\Bzb \to \Dstarp \proton\antiproton\pip$\\$\Dstarp\to\Dz\pip$\\$\Dz\to\Km\pip$\vspace{5 pt}\end{minipage} &
 $ \Bzb_{\mathrm {fake}}\to\pip   \antiproton\textcolor{black}{\biggl[} \proton \Bigl[\bigl[\Km \pip\bigr]_{\Dz}\textcolor{black}{\biggr]_{{{\mathrm{fake}\,}\LCp}}} \pip\Bigr]_{\Dstarp}$  
& $(2.44 \pm 0.12)\times 10^{-3}$ &  $ 13.4$ & 96.9\% & 0.1 \\
\hline\hline
\end{tabular}
        \label{tab:PseudoMinvBG_1}
\end{table*}
\subsubsection{$\Bm\to\SigmaCpl\antiproton\pim$}\label{txt:backgrounds_SigmaCpl}
\begin{figure*}[htpb]
        \centering
          \includegraphics[width=0.95\textwidth]{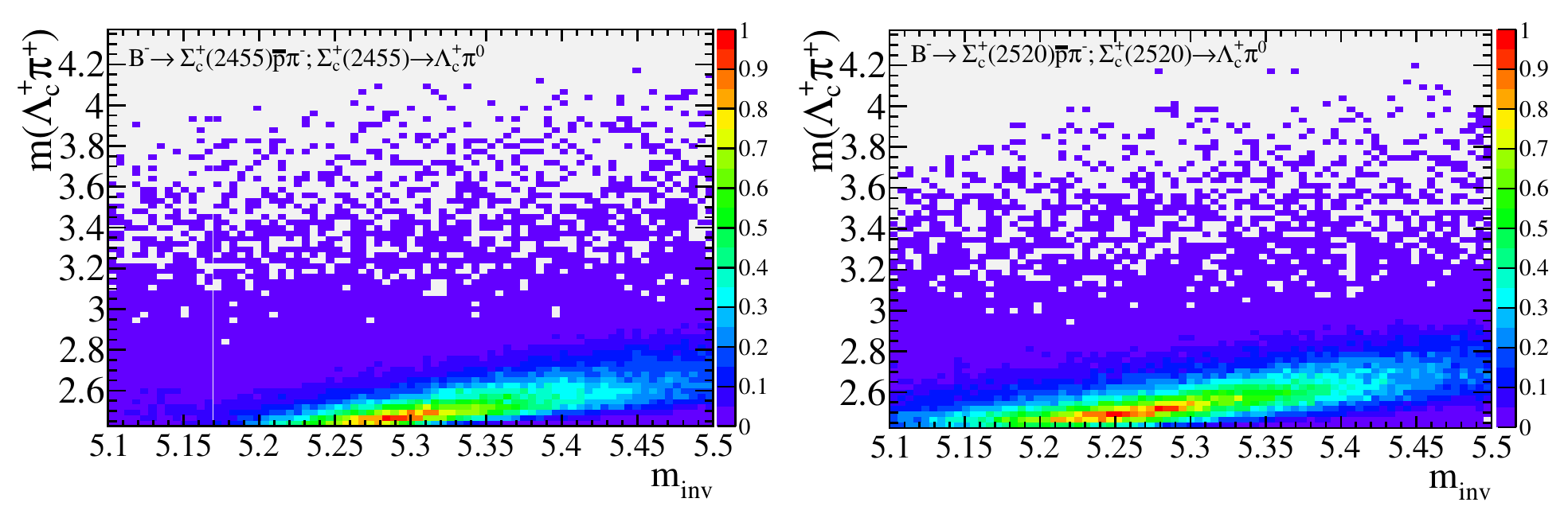}
          \caption{Simulated events with  $\Bm\to\SigmaC(2455)^{+}\antiproton\pim$ (left) and $\Bm\to\SigmaC(2520)^{+}\antiproton\pim$ (right) decays, where $\Sigma_{c}^{{\color{black}{+}}}\to\LCp\pi^{\color{black}{0}}$ decays are reconstructed as $\SigmaC^{+{\color{black}{+}}}\to\LCp\pi^{{\color{black}{+}}}$; these events accumulate in the signal regions of \minv and in \mpp. The MC-generated events are reconstructed as $\Bzb\to\LCp\antiproton\pip\pim$. The color scale indicates the relative contents of a bin compared to the maximally occupied bin (color online).}
          \label{fig:Bm_1}
 \end{figure*}

Events from $\Bm\to\SigmaCpl\antiproton\pim$ decays with $\SigmaC(2455)^{+}\to\LCp\piz$ or $\SigmaC(2520)^{+}\to\LCp\piz$  are found to have a signal-like shape in \minv and \mpp. Because of the low-momentum \piz daughters in the $\SigmaC(2455,2520)^{+}$ center-of-mass systems, fake $\SigmaC(2455,2520)^{++}$ can be generated by replacing the \piz with a \pip from the \Bp. Figure \ref{fig:Bm_1} shows the distributions of MC-generated events. These events cluster in the \minv signal region as well as in \mpp in the $\SigmaC(2455)^{++}$ and $\SigmaC(2520)^{++}$ signal regions. A correlation between \minv and \mpp is apparent. No significant structures are found in MC-generated events with nonresonant $\Bm\to\LCp\antiproton\pim\piz$ or with $\Bm\to\SigmaC(2800)^{+}\antiproton\pim$ events due to the softer momentum constraints on the \piz.
\subsubsection{Combinatorial background with genuine \SigmaC events}\label{txt:backgrounds_CombiSigmaC}
In both MC and data-sideband events, combinatorial background events with genuine $\SigmaC(2455,2520)^{++,0}$ resonances are found to be distributed differently than purely combinatorial background events without \SigmaC  resonances. These events produce a signal-like structure in \mpp or \mpm, but are distributed in \minv similarly to purely combinatorial background. However, since combinatorial background events with genuine $\SigmaC(2455,2520)^{++,0}$ resonances scale differently in \minv than purely combinatorial background events, no simple combined PDF can be constructed. Thus, both combinatorial background sources are treated as separate background classes.
\subsubsection{$\Bzb\to\LCp\antiproton\pip\pim$ events without a \SigmaC signal}\label{txt:backgrounds_FourBodySignalBkg}
Events also appear as background in the \mpppm distribution when they contain decays into the four-body final state $\Bzb\to\LCp\antiproton\pip\pim$, not via the signal \SigmaC resonance. For example, decays such as $\Bzb\to\SigmaCplpl\antiproton\pim$ are distributed as background to $\Bzb\to\SigmaCz\antiproton\pip$ events in \mpm but as signal in \minvn. Therefore, decays to  $\Bzb\to\LCp\antiproton\pip\pim$ not cascading via the signal resonance are included as a background class.
\subsection{Fit Strategy}\label{txt:FitStrategy}
The signal yields of resonant decays are determined in binned maximum-likelihood fits to the two-dimensional distributions $\minvn\,{:}\,\mpp$ and $\minvn\,{:}\,\mpm$. Since background events from  $\Bm\to\SigmaCpl\antiproton\pim$ decays are distributed similarly to signal events in all examined variables, one-dimensional measurements of the signal yield will not suffice. By extracting the signal yield in the $\minvn\,{:}\,\mpppm$ plane, we exploit the fact that the distributions of $\Bm\to\SigmaC(2455,2520)^{+}\antiproton\pim$ events are more correlated in these variables than signal events.
\subsubsection{Type of PDFs}
Signal and background sources are divided into two classes of probability density functions. Background sources without significant correlations between \minv and \mpppm are described with analytical PDFs; independent analytical PDFs are used for each of the two variables and a combined two-dimensional PDF is formed by multiplication of the one-dimensional functions. Signal and background sources with correlations between \minv and \mpppm are described with binned histogram PDFs ${\mathrm{H}_{i}} = S_{i}\cdot{\mathrm{h\comment{ist}}}_{i}\left(\minvn,\mpppm\right)$. For each source, a histogram ${\mathrm{h\comment{ist}}}_{i}\left(\minvn,\mpppm\right)$ is generated from MC events, which takes correlations into account by design. Each histogram ${\mathrm{h\comment{ist}}}_{i}$ is scaled with a parameter $S_{i}$, which is allowed to float in the fit. Histogram PDFs are used for all resonant signal decays and peaking background decays $\Bm\to\SigmaC(2455,2520)^{+}\antiproton\pim$.\\\indent
In the fits to the two-dimensional distributions, the integrals of the analytical PDFs for each bin are calculated. Table \ref{tab:SignalBkg_1} lists the PDFs and indicates whether they are included in the fit to $\minvn\,{:}\,\mpp$ for $\Bzb\to\SigmaCplpl\antiproton\pim$ events or in the fit to $\minvn\,{:}\,\mpm$ for $\Bzb\to\SigmaCz\antiproton\pip$ events.
\subsubsection{Histogram PDF verification}\label{txt:HistoPDFVerification}
When using a histogram PDF in fits, results prove to be sensitive to differences between data events and MC-generated events. As a cross-check, the projections onto \minv are compared between data and MC simulation. The distributions are fitted using a Gaussian function to describe signal events. The means differ between data and MC by  $\Delta=(2.30\pm0.25)\mevcc$. The mass shift does not depend on the \LCp  candidate selection or on \mpppm. The most probable explanation for the difference is an underestimation of the SVT material in the simulation, as studied in detail in Ref. \cite{Lit:LambdaCMassMeasurement}. Baryonic decays are especially affected by this issue, since heavier particles such as protons suffer more from such an underestimation compared to lighter particles. In each MC event, the baryon momenta $\left|\vec{{p}}_{p_{\LCp}}\right|$ and $\left|\vec{{p}}_{p_{\Bzb}}\right|$ are therefore increased by $2.30\mevc$ and the particle energy is adjusted accordingly.\\\indent
In the \mpppm distributions, the means of the masses of the $\SigmaC(2455)^{++,0}$ baryons differ between data and MC by $\Delta{m}=\left(0.441\pm0.095\right)\mevcc$. This effect originates from outdated $\SigmaC(2455)$ mass inputs in the MC generation, and so this shift is not covered by the correction for detector density. $\SigmaC(2455)$ events are especially sensitive to such mass differences due to their narrow width. The effect is taken into account by shifting each MC event in \mpppm by $+0.441\gevcc$. The fully corrected data sets are used to generate the histogram PDFs employed in the fits to data.
\subsubsection{Combinatorial background PDF}
The combinatorial background is described by the PDF ${\mathrm{BG}}_{\mathrm{Combi\, Bkg}}$ term given in Table \ref{tab:SignalBkg_1}. It consists of two separable functions: for \minv we use a first-order Chebyshev polynomial ${{\mathrm{X}}_{\mathrm{Combi\, Bkg\,w\,\SigmaC}}^{\mathrm{Chebyshev}}}\left(\minvn;b\right)$ with a slope parameter $b$ and for \mpppm a phenomenological function,
\begin{align}
\label{math:2TschebycheffPoly_CombiBkg_1}
& \mathrm{Y}_{\mathrm{Combi\, Bkg}}\left(\mpppm;p,q;e_{\mathrm{up}},e_{\mathrm{low}}\right) =  \\
& \left(c - \mpppm\right)^{p} \cdot \left(\mpppm-e_{\mathrm{low}}\right)^{q} \cdot e_{\mathrm{up}}\,.\nonumber
\end{align}
The upper and lower phase-space boundaries in \mpppm are constants $e_{\mathrm{low}}= 2.4249\gevcc$ and $e_{\mathrm{up}}=4.215\gevcc$. The phenomenological constant $c = 4.108\gevcc$ is obtained from MC and, for estimating an systematic uncertainty, varied within the values found in MC.
The exponent terms $p$ and $q$  are allowed to float in the fits to MC and data. In Table \ref{tab:SignalBkg_1}, $S_{\mathrm{Combi\, Bkg}}$ is the overall scaling parameter of the combinatorial background PDF.
\subsubsection{Combinatorial background with genuine \SigmaC PDF}
Combinatorial background events with genuine \SigmaC resonances are described by uncorrelated functions in \minv and \mpppm. A first-order Chebyshev polynomial ${{\mathrm{X}}_{\mathrm{Combi\, Bkg\,w\,\SigmaC}}^{\mathrm{Chebyshev}}}\left(\minvn;b\right)$  in \minv is multiplied by a nonrelativistic Breit-Wigner function in \mpppm,
\begin{align}{\mathrm{BW}}_{\mathrm{Combi\, Bkg\,w\,\SigmaC}}\left(\mpppm;\mu,\Gamma\right)=\\
\frac{1}{\pi \cdot \left[\left(\mpppm-\mu\right)^{2} + \left(\frac{\Gamma}{2}\right)^{2}\right]},\nonumber
\end{align}
with mean $\mu$, width $\Gamma$, and an overall scaling factor $S_{{\mathrm{Combi\,Bkg\,w\,\SigmaC}}}$, to form a two-dimensional PDF (${\mathrm{BG}}_{{\mathrm{Combi\,Bkg\,w\,\SigmaC}}}$ in Table \ref{tab:SignalBkg_1}) in \mpppm.\\\indent
The PDFs for combinatorial background with and without genuine \SigmaC resonances are validated using studies with MC events and from fits to data within the \minv sidebands of the $\minvn\,{:}\,\mpppm$ planes.
\subsubsection{non-\SigmaC $\Bzb\to\LCp\antiproton\pip\pim$}
Events with $\Bzb\to\LCp\antiproton\pip\pim$ decays but  without signal \SigmaC resonances are described by the product of a phenomenological function in \mpppm,
\begin{align} 
\label{math:2TschebycheffPoly_CombiBkg_Total_1}
& \mathrm{Y}_{\mathrm{non-\SigmaC}}(\mpppm;n,p,q,r;e_{\mathrm{up}}) =  \\
& (\mpppm - r) \cdot \left[\mpppm\right]^{p} \cdot \left(e_{\mathrm{up}}-\mpppm\right)^{q},\nonumber
\end{align}
and a Gaussian function  $\mathrm{X}_{\mathrm{non-\SigmaC}}\left(\minvn;\mu,{\sigma}\right)$ in \minvn.
Fits to mixtures of MC samples (denoted as {\it toy MC mixtures}) are used to validate the combined two-dimensional PDF (${\mathrm{BG}}_{{\mathrm{non-\SigmaC}}}^{\mathrm{Gauss}}$ in Table \ref{tab:SignalBkg_1}). This procedure is designed to take into account the fact that $\Bzb\to\LCp\antiproton\pip\pim$ decays without signal \SigmaC resonances can proceed via various other intermediate states besides direct decays into the four-body final state. Since the true composition of $\Bzb\to\LCp\antiproton\pip\pim$ decays without signal \SigmaC resonances is unknown, toy MC sets are created by combining randomly chosen numbers of MC events with completely nonresonant signal decays, signal decays with intermediate non-\SigmaC resonances such as $N^{*}$ or $\rho$, and signal decays with non-signal \SigmaC resonances.\\\indent
In the fits to toy MC samples, a quadratic dependency on \mpppm of the signal-Gaussian width in \minv is observed. This is taken into account by parameterizing the width as ${\sigma}\left(\mpppm;a_{\sigma},b_{\sigma},c_{\sigma}\right) = c_{\sigma} \cdot \left[a_{\sigma}\cdot{\mpppm}^{2} +  b_{\sigma}\cdot \mpppm +  1\right]$. In fits to data, the width parameters are fixed to the values obtained from fits to MC; a systematic uncertainty on the shape is included by varying the parameters within the parameter range obtained from fits to toy MC.\\\indent
Global PDFs for fits to $\minvn\,{:}\,\mpp$ and to $\minvn\,{:}\,\mpm$ are formed from sums over the signal and background PDFs as listed in Table \ref{tab:SignalBkg_1}. The global PDFs are validated on toy MC samples of randomly chosen numbers of events from the signal and background classes.
\begin{table*}[htpb]
  \centering
\caption{The PDF types for signal and background sources as defined in Sec. \ref{txt:FitStrategy} (see text for details). In the second column, $S_{i}$ denote scaling factors, ${\mathrm{h}}_{i}$ histograms, and ${\mathrm{X}}_{i}$, ${\mathrm{Y}}_{i}$, ${\mathrm{BW}}_{i}$ analytical functions as described in the text. The third and fourth columns indicate in which global fit  to the planes  $\minvn\,{:}\,\mpp$ or $\minvn\,{:}\,\mpm$ a particular PDF is included.\vrule width0pt depth12pt}
\begin{tabular}{l c c c}\hline\hline
{ Mode} & { PDF} & { $\minvn\,{:}\,\mpp$} & { $\minvn\,{:}\,\mpm$}\\\hline
$\Bzb\to\SigmaC(2455)^{++}\antiproton\pim$ &  ${\mathrm{H}}_{\mathrm{\SigmaC(2455)^{++}}} =S_{\SigmaC(2455)^{++}} \cdot {\mathrm{h\comment{ist}}}_{\SigmaC(2455)^{++}}$ & $\checkmark$ & \\[3 pt]
$\Bzb\to\SigmaC(2520)^{++}\antiproton\pim$&  ${\mathrm{H}}_{\mathrm{\SigmaC(2520)^{++}}} =S_{\SigmaC(2520)^{++}} \cdot {\mathrm{h\comment{ist}}}_{\SigmaC(2520)^{++}}$ & $\checkmark$ &  \\[3 pt]
$\Bzb\to\SigmaC(2455)^{0}\antiproton\pip$ &  ${\mathrm{H}}_{\mathrm{\SigmaC(2455)^{0}}} =S_{\SigmaC(2455)^{0}} \cdot {\mathrm{h\comment{ist}}}_{\SigmaC(2455)^{0}}$ &  & $\checkmark$ \\[3 pt]
$\Bzb\to\SigmaC(2520)^{0}\antiproton\pip$ &  ${\mathrm{H}}_{\mathrm{\SigmaC(2520)^{0}}} = S_{\SigmaC(2520)^{0}} \cdot {\mathrm{h\comment{ist}}}_{\SigmaC(2520)^{0}}$ &  & $\checkmark$ \\[3 pt]
$\Bm\to\SigmaC(2455)^{+}\antiproton\pim$ &  ${\mathrm{BG}}_{\mathrm{\SigmaC(2455)^{+}}} = S_{\SigmaC(2455)^{+}} \cdot {\mathrm{h\comment{ist}}}_{\SigmaC(2455)^{+}}$ & $\checkmark$ & \\[3 pt]
$\Bm\to\SigmaC(2520)^{+}\antiproton\pim$ &  ${\mathrm{BG}}_{\mathrm{\SigmaC(2520)^{+}}} = S_{\SigmaC(2520)^{+}} \cdot {\mathrm{h\comment{ist}}}_{\SigmaC(2520)^{+}})$ & $\checkmark$ & \\[3 pt]
Combinatorial background & {\begin{minipage}[H]{0.45\textwidth}{
\begin{align}
{\mathrm{BG}}_{\mathrm{Combi\, Bkg}} & = S_{\mathrm{Combi\, Bkg}}   \nonumber \\
    & \times \mathrm{Y}_{\mathrm{Combi\, Bkg}}\left(\mpppm;p,q;e_{\mathrm{up}},e_{\mathrm{low}}\right)  \nonumber \\
    & \times {\mathrm{X}}_{\mathrm{Combi\, Bkg}}^{\mathrm{Chebyshev}}\left(\minvn;b\right)   \nonumber
\end{align}}\end{minipage}} & $\checkmark$ & $\checkmark$  \\[3 pt]
\begin{minipage}[H]{0.25\textwidth}Combinatorial background\newline with genuine \SigmaC\end{minipage} & {\begin{minipage}[H]{0.45\textwidth}{\begin{align}
\label{txt:Def:CombinatorialBkgSigmaC2D_1}
{\mathrm{BG}}_{{\mathrm{Combi\,Bkg\,w\,\SigmaC}}} & = S_{\mathrm{Combi\, Bkg\,w\,\SigmaC}} \nonumber\\
 &  \times {\mathrm{BW}}_{\mathrm{Combi\, Bkg\,w\,\SigmaC}}\left(\mpppm;\mu,\Gamma\right)  \nonumber\\
 &  \times {{\mathrm{X}}_{\mathrm{Combi\, Bkg\,w\,\SigmaC}}^{\mathrm{Chebyshev}}}\left(\minvn;b\right) \nonumber
\end{align}}\end{minipage}} & $\checkmark$ & $\checkmark$  \\[3 pt]
non-\SigmaC $\Bz\to\LCp\antiproton\pip\pim$ & {\begin{minipage}[H]{0.45\textwidth}{\begin{align}
 {\mathrm{BG}}_{{\mathrm{non-\SigmaC}}} & = S_{{\mathrm{non-\SigmaC}}} \nonumber\\
 &  \times  \mathrm{Y}_{{\mathrm{non-\SigmaC}}}(\mpppm;p,q,r;e_{\mathrm{psb}}) \nonumber \\
 &\times  \mathrm{X}_{\mathrm{non-\SigmaC}}^{\mathrm{Gauss}}\left(\minvn;\mu,{\sigma}\left[\mpppm;a_{\sigma},b_{\sigma},c_{\sigma}\right]\right) \nonumber
\end{align}}\end{minipage}} & $\checkmark$ & $\checkmark$ \\[3 pt]
\hline\hline
\end{tabular}
        \label{tab:SignalBkg_1}
\end{table*}
\subsection{Fit Results}
\begin{figure*}[htpb]
\begin{tabular}{ccc}
   \begin{minipage}[H]{0.475\textwidth}
     \centering
     \includegraphics[width=1.0\textwidth]{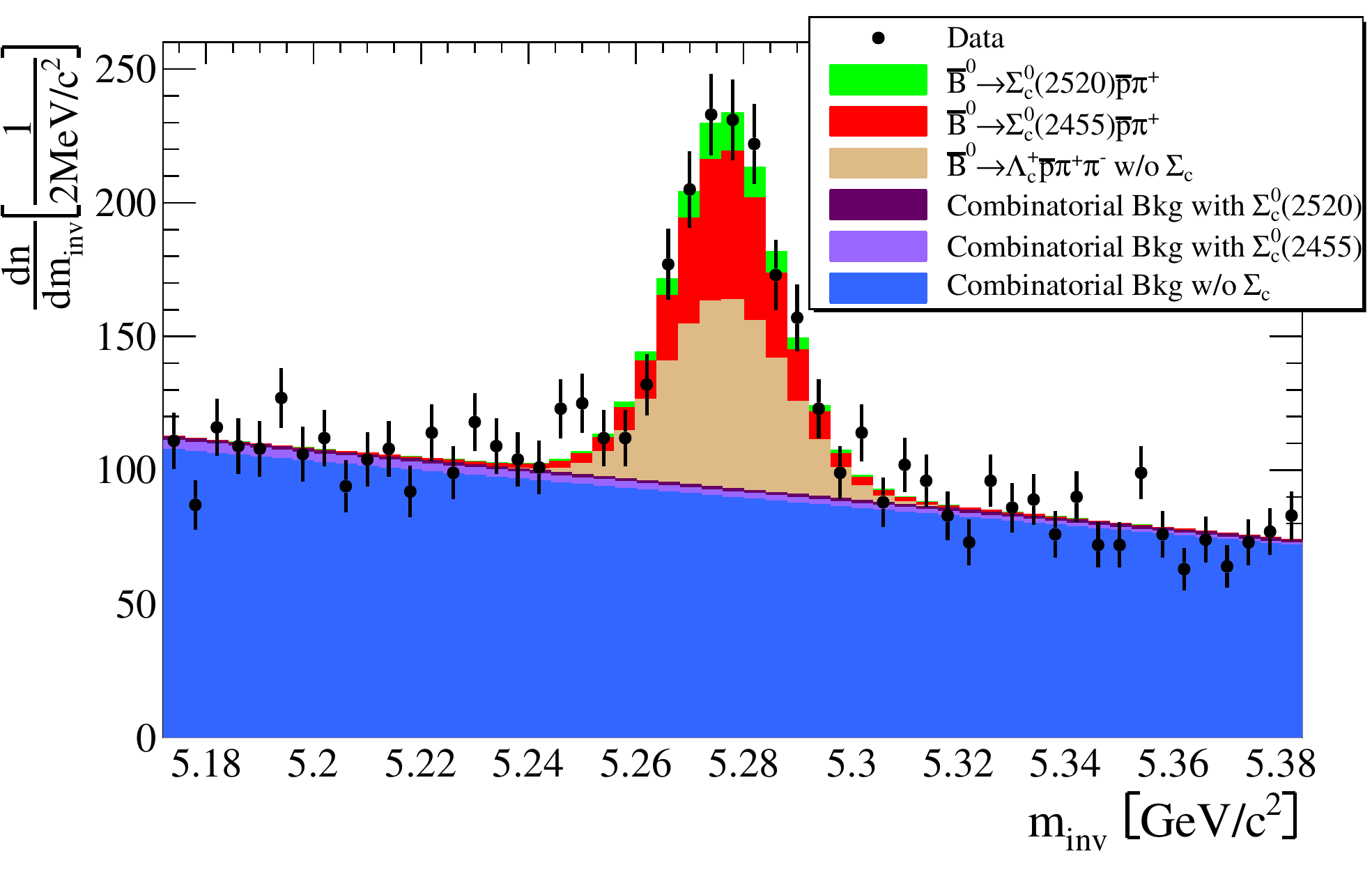}
     \caption{Result of the fit to the $\minvn\,{:}\,\mpm$ plane projected onto the \minv axis. The data are shown as points with error bars; the fitted signal and background PDFs are overlaid as stacked histograms.}
     \label{fig:FitSigmaC0_1}
   \end{minipage}
&
   \begin{minipage}[H]{0.05\textwidth}
\quad
   \end{minipage}

&
        \centering
   \begin{minipage}[H]{0.475\textwidth}
          \includegraphics[width=1.0\textwidth]{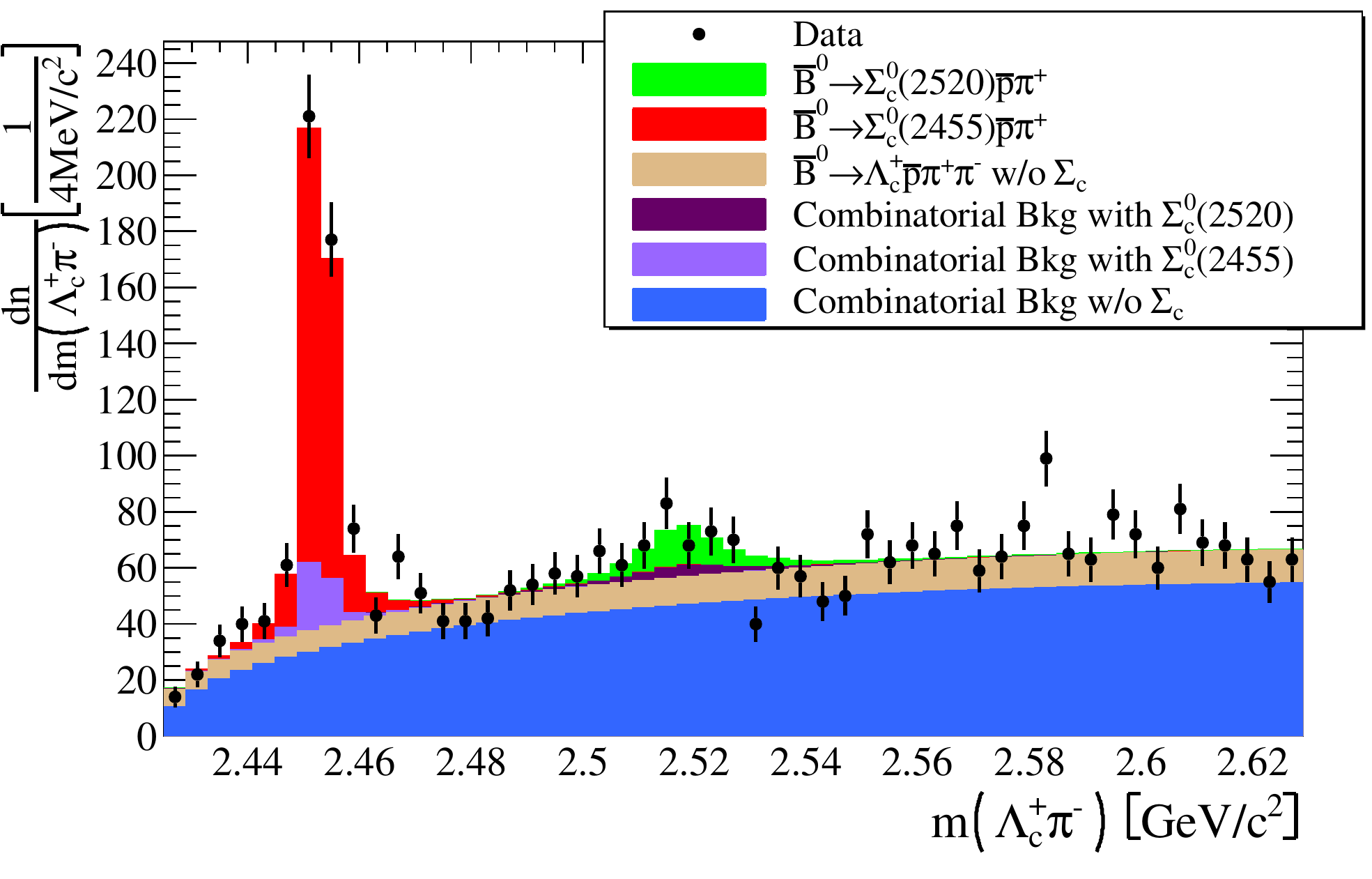}
          \caption{Result of the fit to the $\minvn\,{:}\,\mpm$ plane projected onto the \mpm axis. The data are shown as points with error bars; the fitted signal and background PDFs are overlaid as stacked histograms.}
          \label{fig:FitSigmaC0_2}
   \end{minipage}
\end{tabular}
\end{figure*}

\begin{figure*}[htpb]
\begin{tabular}{ccc}
   \begin{minipage}[H]{0.475\textwidth}
        \centering
          \includegraphics[width=1.0\textwidth]{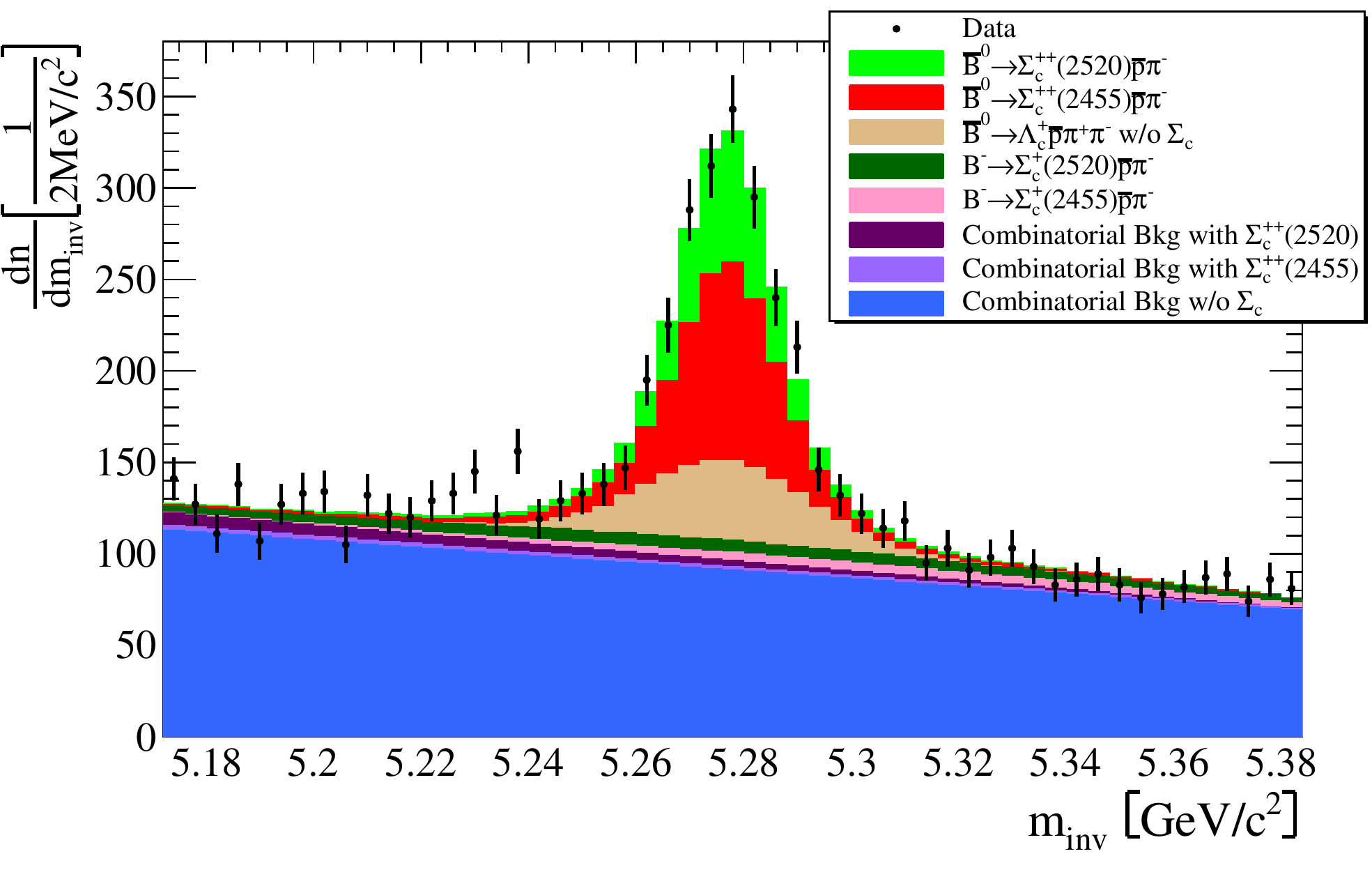}
          \caption{Result of the fit to the $\minvn\,{:}\,\mpp$ plane projected onto the \minv axis. The  data are shown as points with error bars; the fitted signal and background PDFs are overlaid as stacked histograms.}
          \label{fig:FitSigmaCplpl_1}
   \end{minipage}
&
   \begin{minipage}[H]{0.05\textwidth}
\quad
   \end{minipage}
&
   \begin{minipage}[H]{0.475\textwidth}
        \centering
          \includegraphics[width=1.0\textwidth]{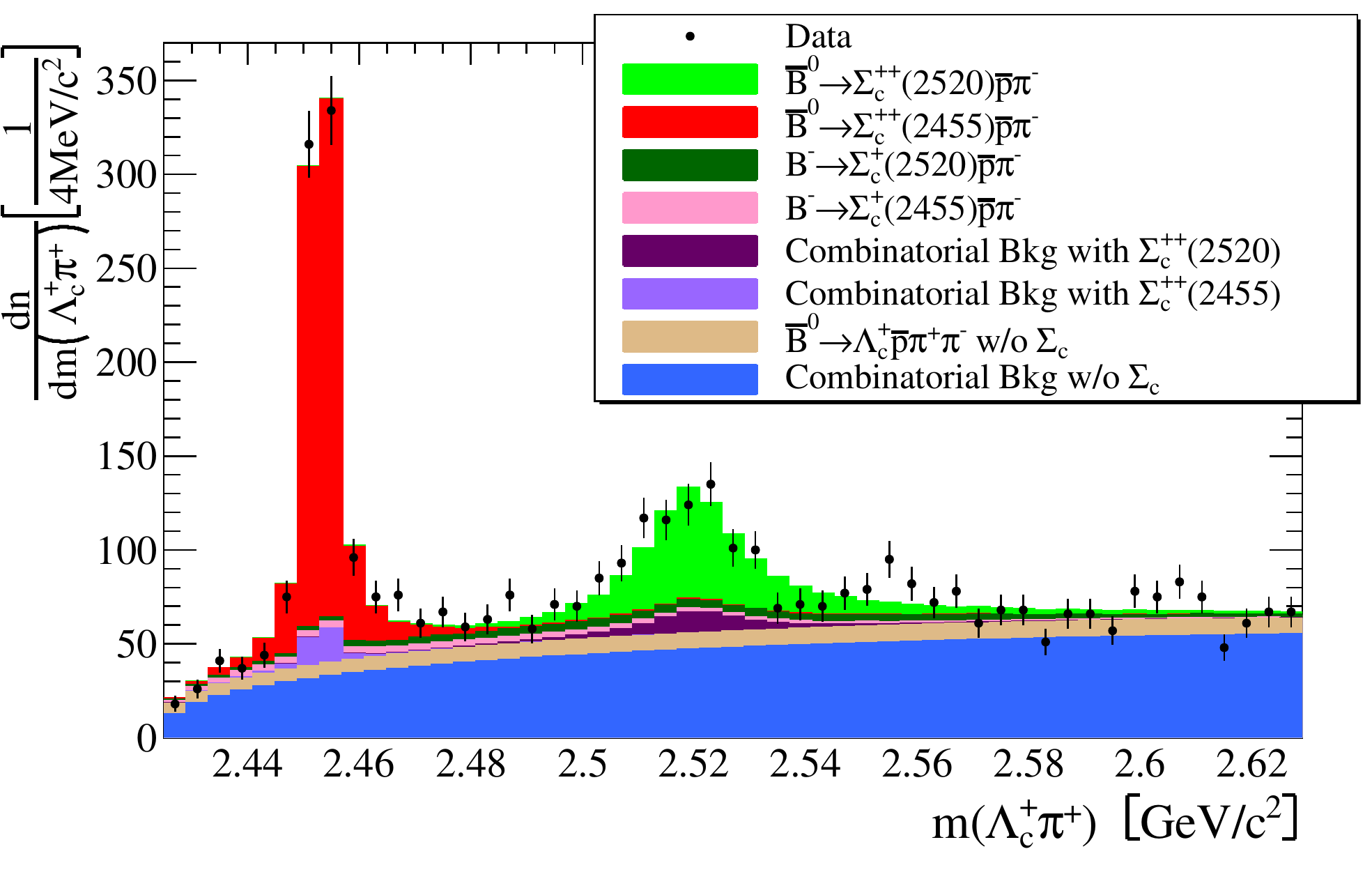}
          \caption{Result of the fit to the $\minvn\,{:}\,\mpp$ plane projected onto the \mpp axis. The data are shown as points with error bars; the fitted signal and background PDFs are overlaid as stacked histograms.}
          \label{fig:FitSigmaCplpl_2}
   \end{minipage}
\end{tabular}
 \end{figure*}

Maximum likelihood fits to data distributions in the $\minvn\,{:}\,\mpp$ and $\minvn\,{:}\,\mpm$ planes are performed in the range $\minvn\in[5.17,5.38]\gevcc$ and $\mpppm\in[2.425,2.625]\gevcc$, covering the regions of $\SigmaC(2455)$ and $\SigmaC(2520)$ resonances.\footnote{Larger \mpppm masses are omitted because of uncertainties on the $\SigmaC(2800)$ mass values. In a study of the related decay $\Bm\to\LCp\antiproton\pim$ \cite{Babar_Stephanie}, a significant difference in the mass is measured for the $\SigmaC(2800)^{0}$ resonances with $m\left(\SigmaC(2800)^{0}\right)=\left(2.846\pm 0.008\right)\gevcc$ compared to the world averaged mass $m\left(\SigmaC(2800)^{0}\right)_{\mathrm{PDG}}=\left(2.802^{+0.004}_{-0.007}\right)\gevcc$ \cite{PDG2010}. Thus, the histogram PDF approach based on MC simulations is not feasible for \SigmaC(2800), since the input mass value is necessary for the MC generation.\label{footnote_sigmac2800}}\\\indent
The fit to $\minvn\,{:}\,\mpm$ converges with $\chiq/{\mathrm{ndf}}=2807/2697$. Figure \ref{fig:FitSigmaC0_1} shows the projection of the two-dimensional fit onto the \minv axis. The fitted PDFs are shown as stacked histograms and are overlayed with the distribution in data. The projection onto the \mpm axis is shown in Fig.\;\ref{fig:FitSigmaC0_2}.\\\indent
In the $\minvn\,{:}\,\mpp$ plane, the fit converges with $\chiq/{\mathrm{ndf}}=2592/2695$. The two-dimensional fit results are shown in Fig.\;\ref{fig:FitSigmaCplpl_1} for the projection onto \minv and in Fig.\;\ref{fig:FitSigmaCplpl_2} for the projection onto \mpp.\\\indent
The measured signal yields are given in Table \ref{tab:FitYields_1}.
\begin{table}[htpb]
  \centering
\caption{Signal yields without efficiency correction from the fits to the $\minvn\,{:}\,\mpppm$ planes. The uncertainties are statistical. \vrule width0pt depth12pt}
\begin{tabular}{l c}\hline\hline\\[-5pt]
{ Mode} & { Signal yield}\\\hline\\[-5pt]
$\Bzb\to\SigmaC(2455)^{0}\antiproton\pip$ & $ 347 \pm 24$ \\
$\Bzb\to\SigmaC(2520)^{0}\antiproton\pip$ & $ 87\pm 27$ \\[5 pt]
$\Bzb\to\SigmaC(2455)^{++}\antiproton\pim$ & $ 723 \pm 32$ \\
$\Bzb\to\SigmaC(2520)^{++}\antiproton\pim$ & $458 \pm 38$\\[5 pt]
$\Bm\to\SigmaC(2455)^{+}\antiproton\pim$ & $ 164 \pm 104$ \\
$\Bm\to\SigmaC(2520)^{+}\antiproton\pim$ & $ 273 \pm 133$\\\hline\hline
\end{tabular}
        \label{tab:FitYields_1}
\end{table}

\subsection{Signal event distributions} 
\begin{figure}[htpb]
  \begin{minipage}[H]{0.45\textwidth}
    \centering
    \subfigure{
      \includegraphics[width=0.95\textwidth]{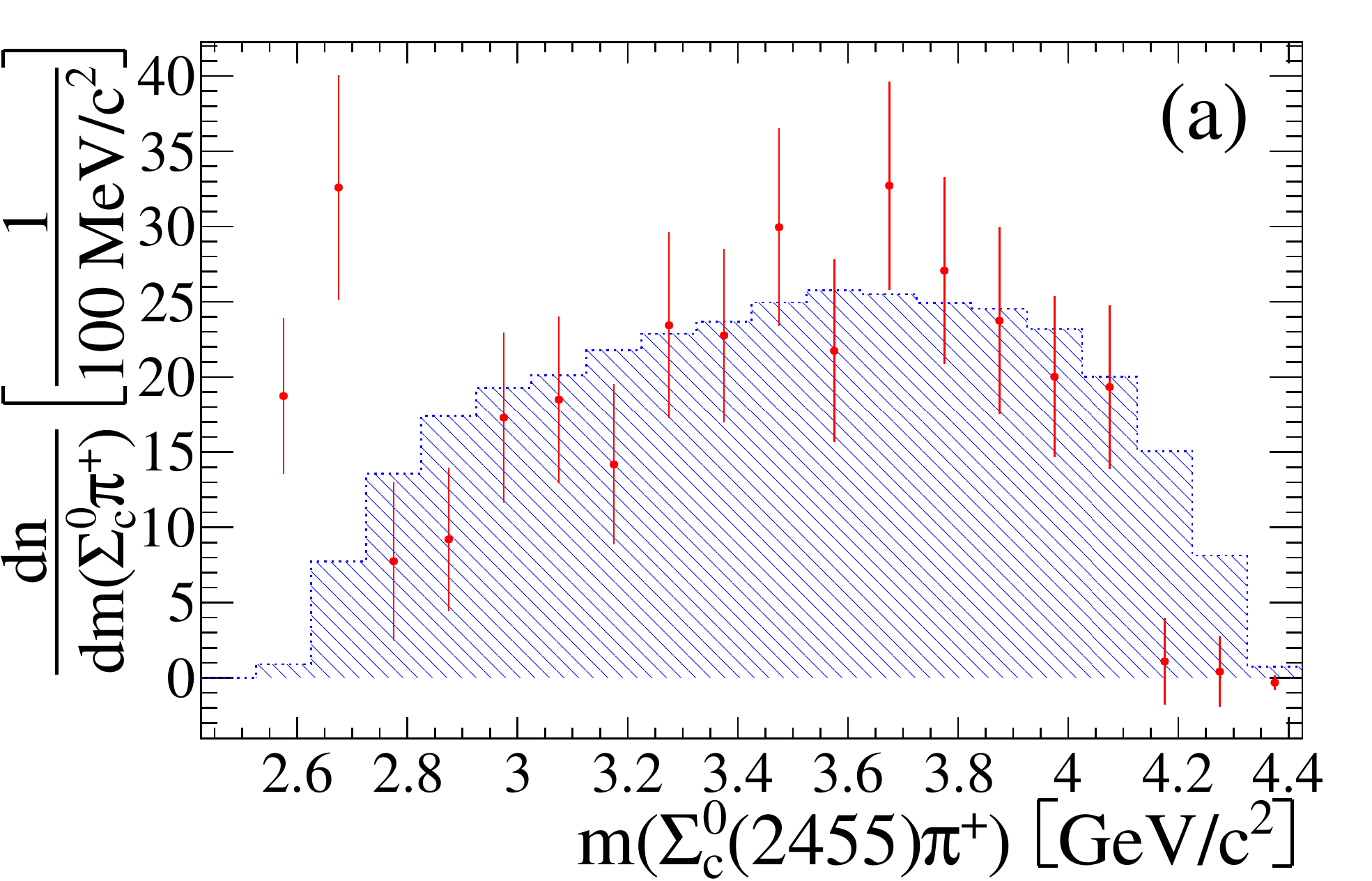}
    \label{fig:SplotsSigmaC02455Vgla}
    }
    \subfigure{
      \includegraphics[width=0.95\textwidth]{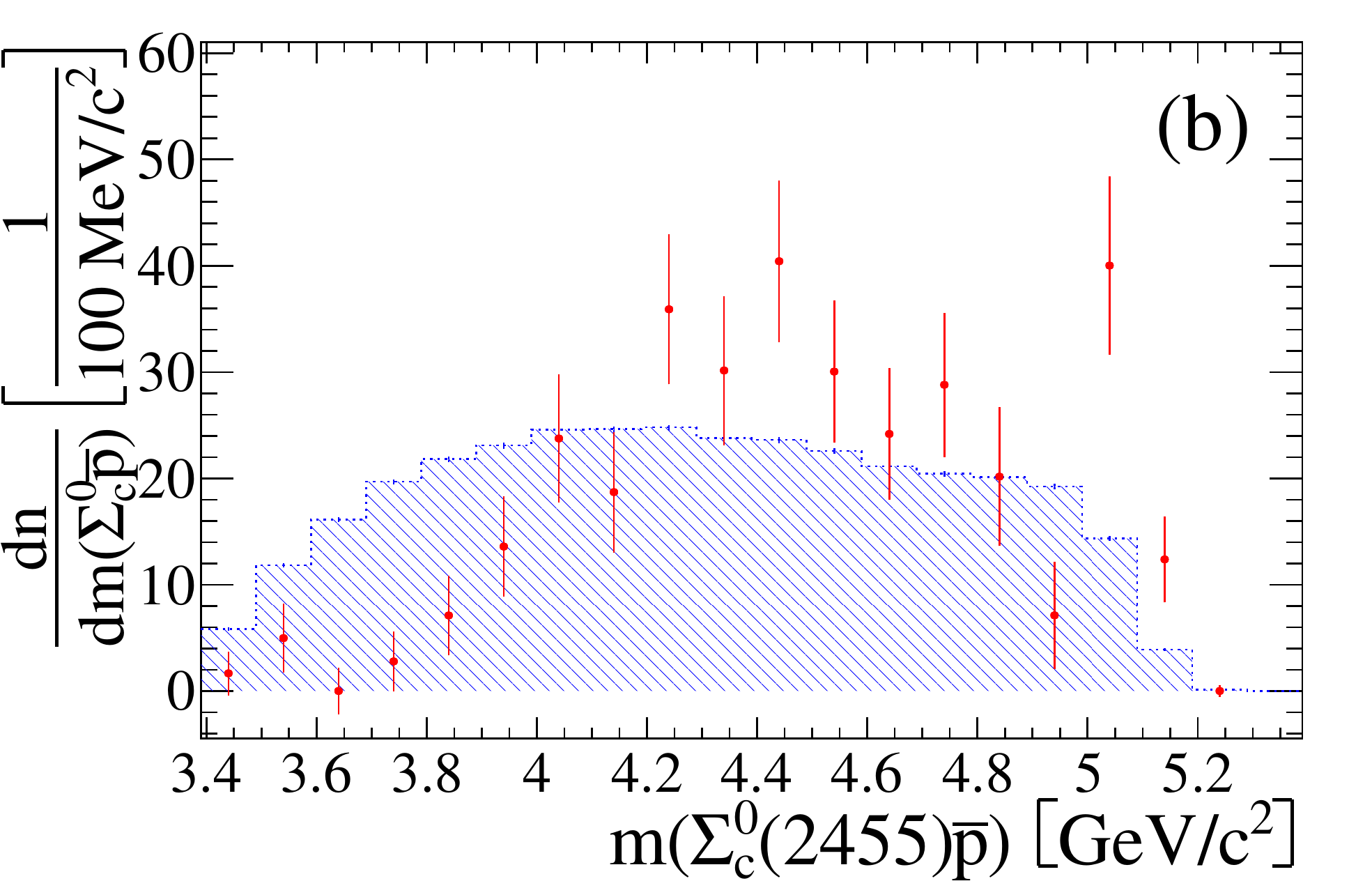}
    \label{fig:SplotsSigmaC02455Vglb}
    }
    \subfigure{
      \includegraphics[width=0.95\textwidth]{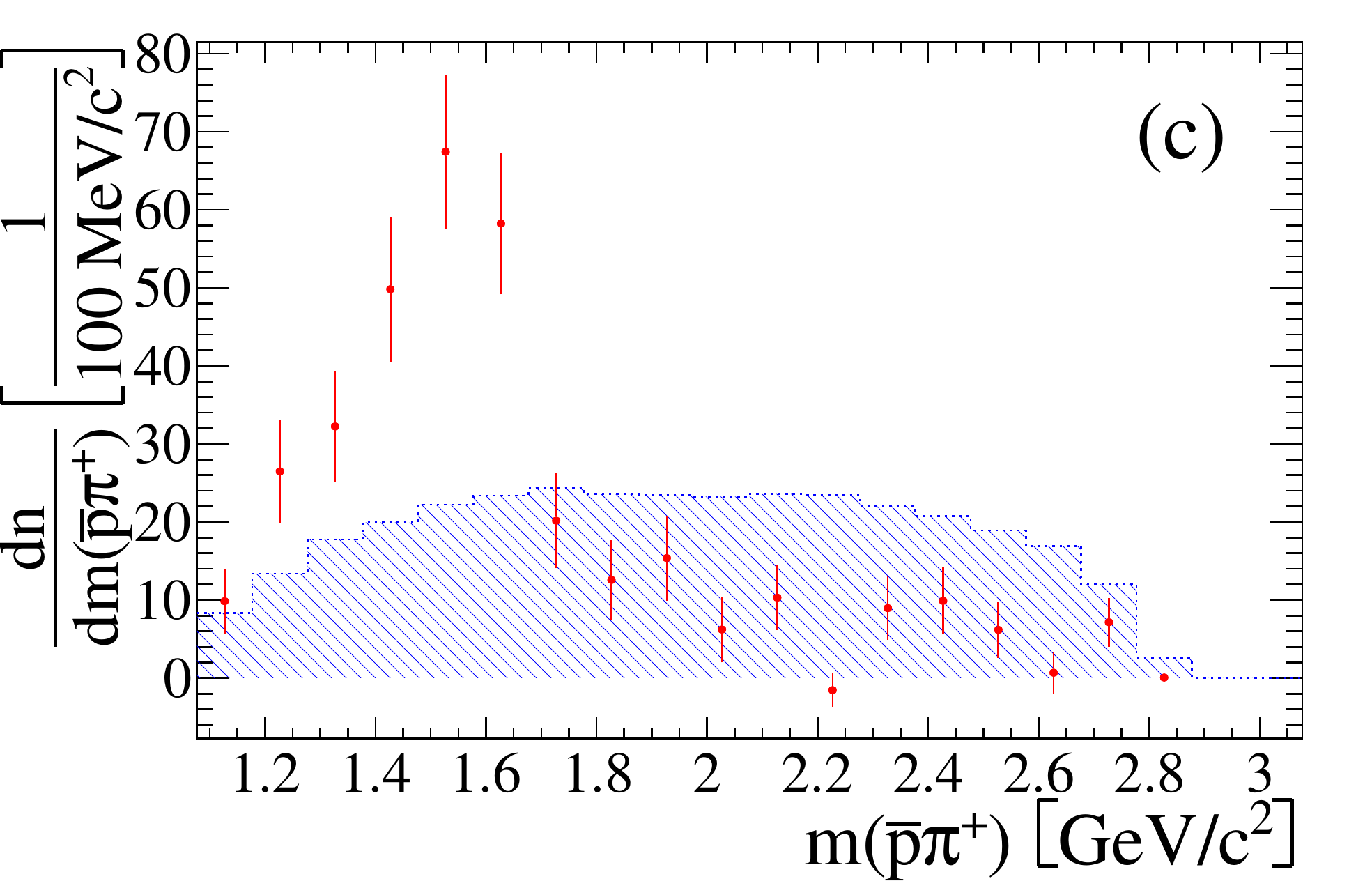}
    \label{fig:SplotsSigmaC02455Vglc}
    }
    \caption{Invariant mass distributions from $\Bzb\to\SigmaC(2455)^{0}\antiproton\pip$ signal events extracted with the \sPlot method. 
Data points are displayed in comparison with the distribution of reconstructed phase-space generated $\Bzb\to\SigmaC(2455)^{0}\antiproton\pip$  MC events scaled to the same number of entries (shaded histograms).}
    \label{fig:SplotsSigmaC02455Vgl}
\end{minipage}
\end{figure}

\begin{figure}[htpb]
\begin{minipage}[H]{0.45\textwidth}
    \centering
    \subfigure{
      \includegraphics[width=0.95\textwidth]{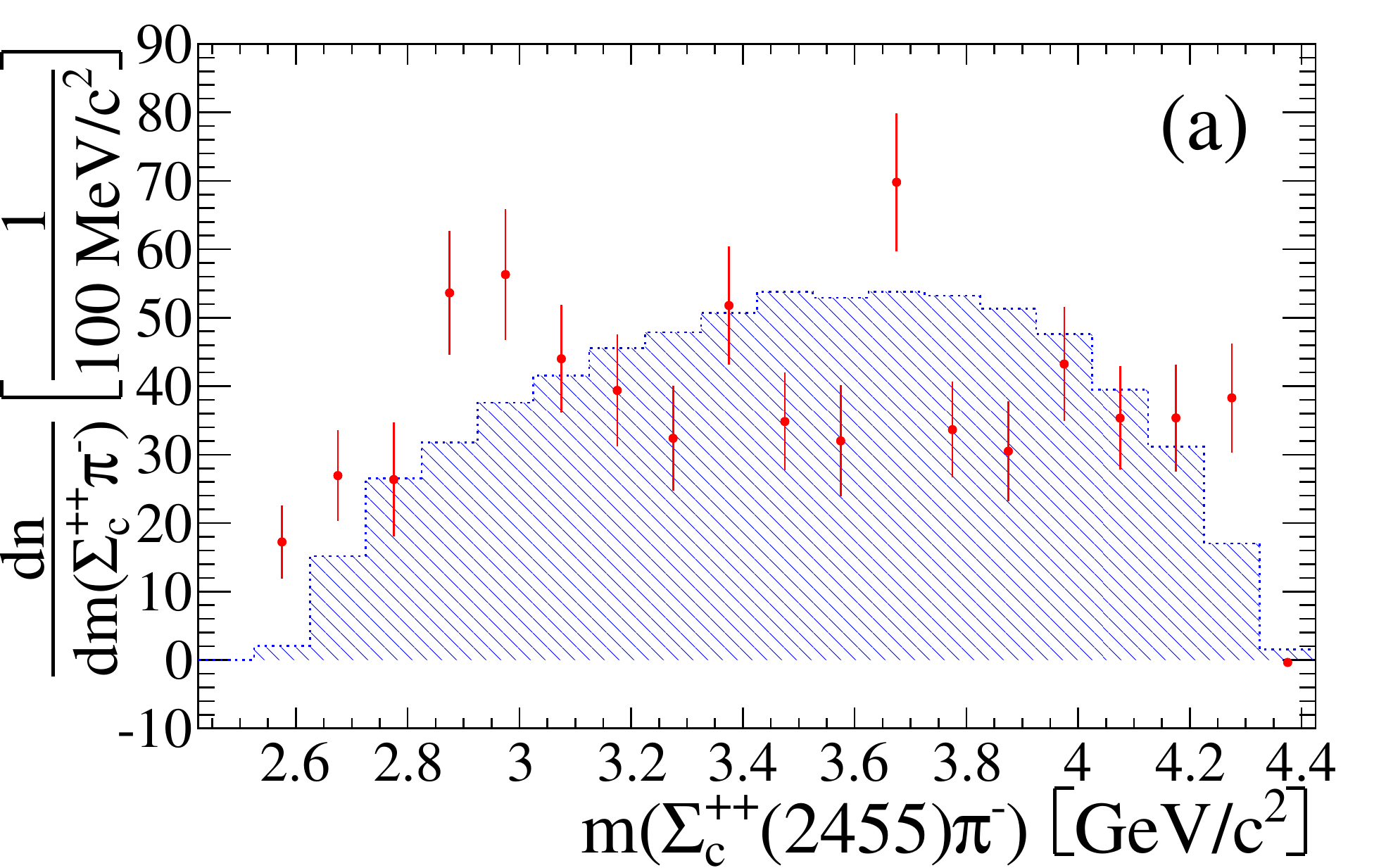}
      \label{fig:SplotsSigmaCplpl2455Vgla}
    }
    \subfigure{
      \includegraphics[width=0.95\textwidth]{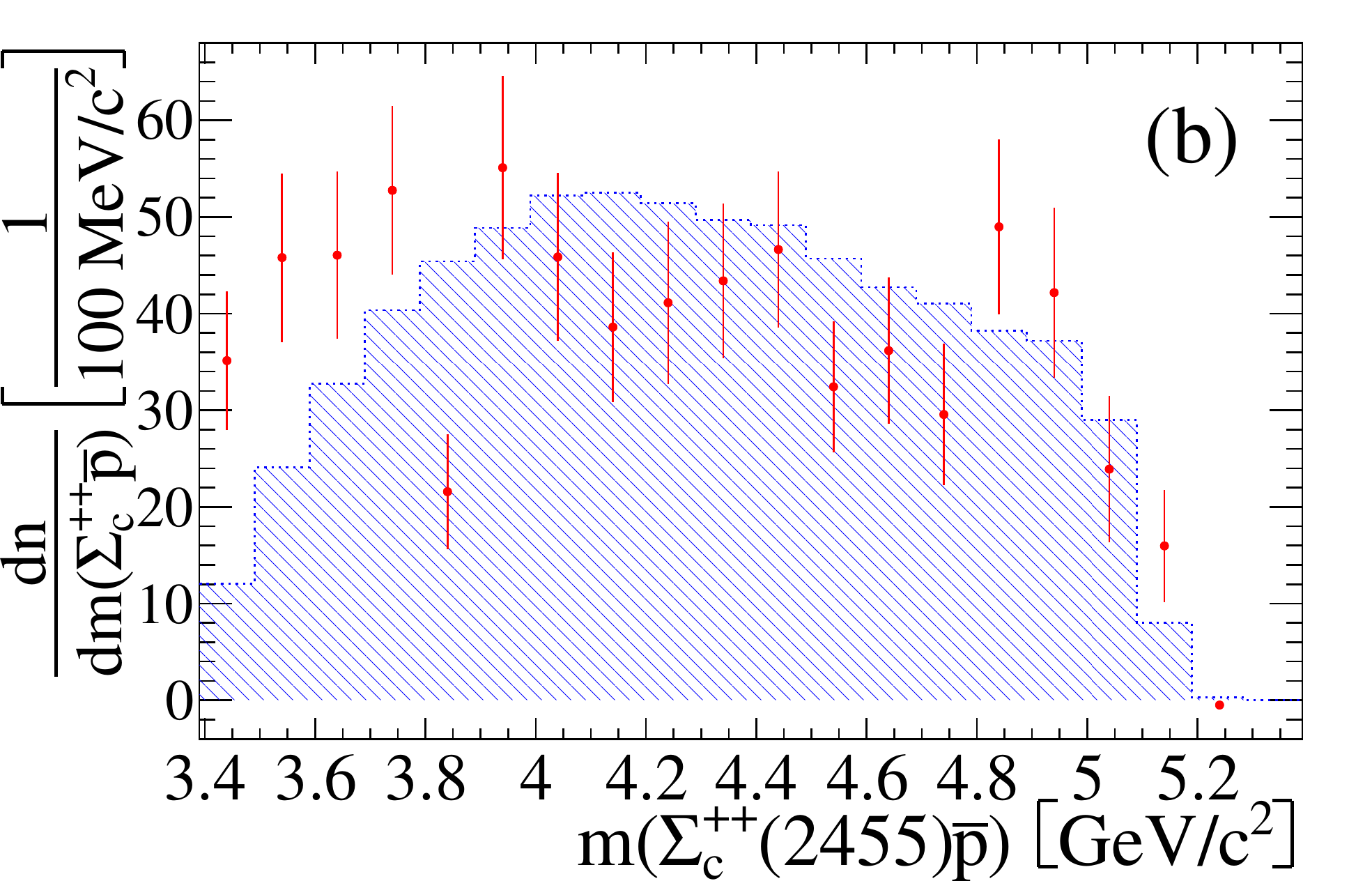}
    \label{fig:SplotsSigmaCplpl2455Vglb}
    }
    \subfigure{
      \includegraphics[width=0.95\textwidth]{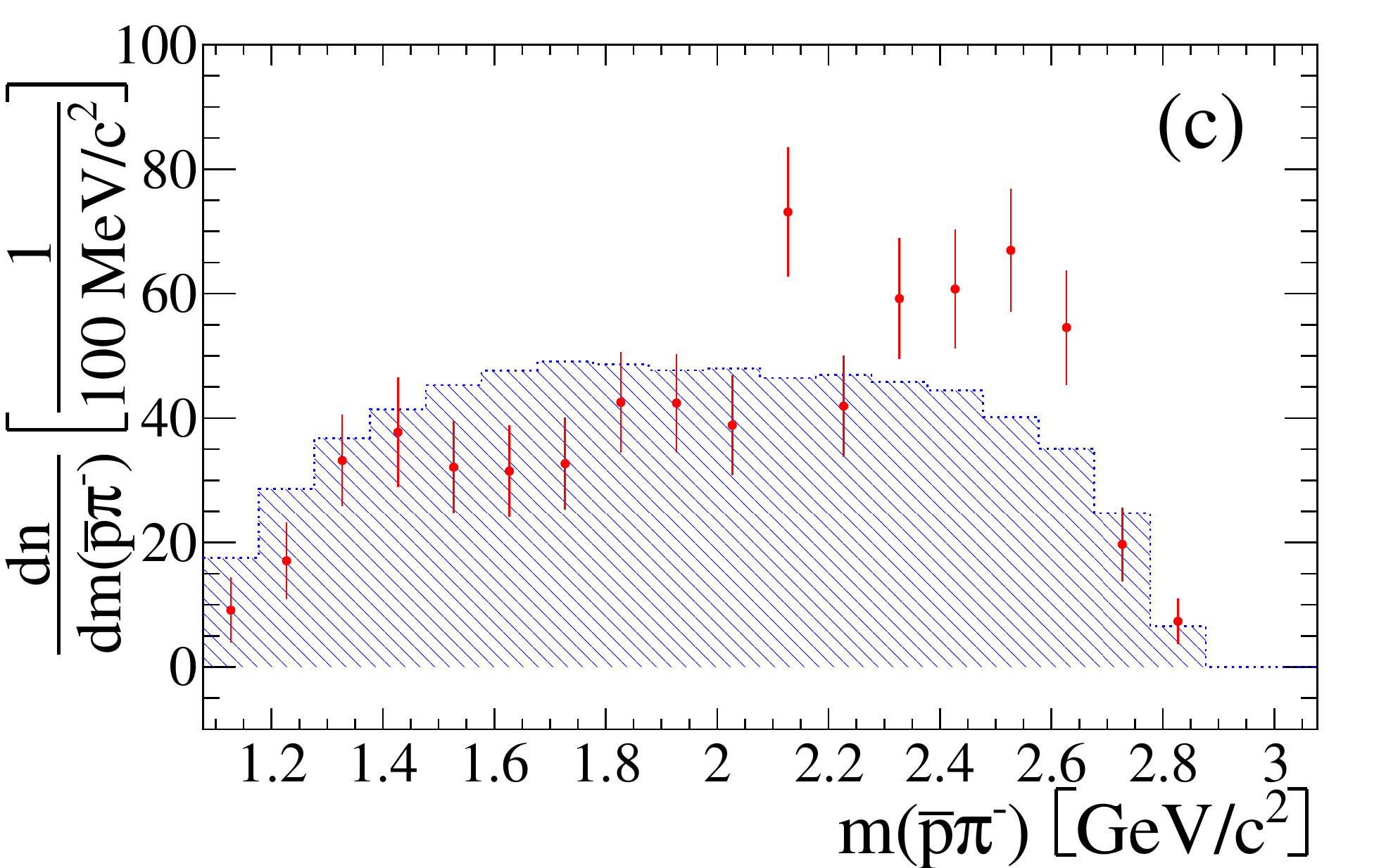}
    \label{fig:SplotsSigmaCplpl2455Vglc}
    }
    \caption{Invariant mass distributions from $\Bzb\to\SigmaC(2455)^{++}\antiproton\pim$ signal events extracted with the \sPlot method. 
Data points are displayed in comparison with the distribution of reconstructed phase-space generated $\Bzb\to\SigmaC(2455)^{++}\antiproton\pim$  MC events scaled to the same number of entries (shaded histograms).}
    \label{fig:SplotsSigmaCplpl2455Vgl}
\end{minipage}
\end{figure}

\begin{figure}[htpb]
  \begin{minipage}[H]{0.45\textwidth}
    \centering
    \subfigure{
      \includegraphics[width=0.95\textwidth]{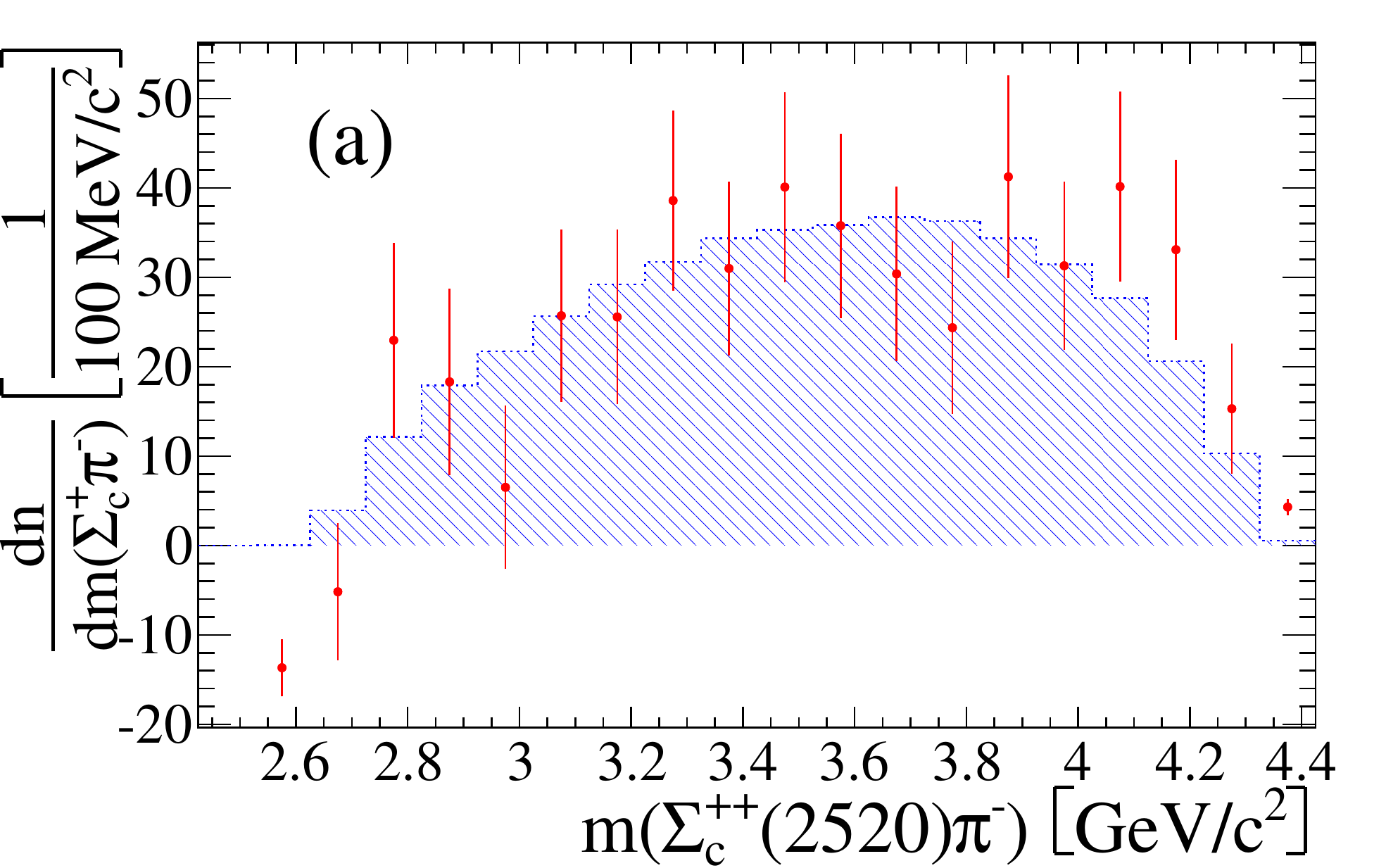}
      \label{fig:SplotsSigmaCplpl2520Vgla}
    }
    \subfigure{
      \includegraphics[width=0.95\textwidth]{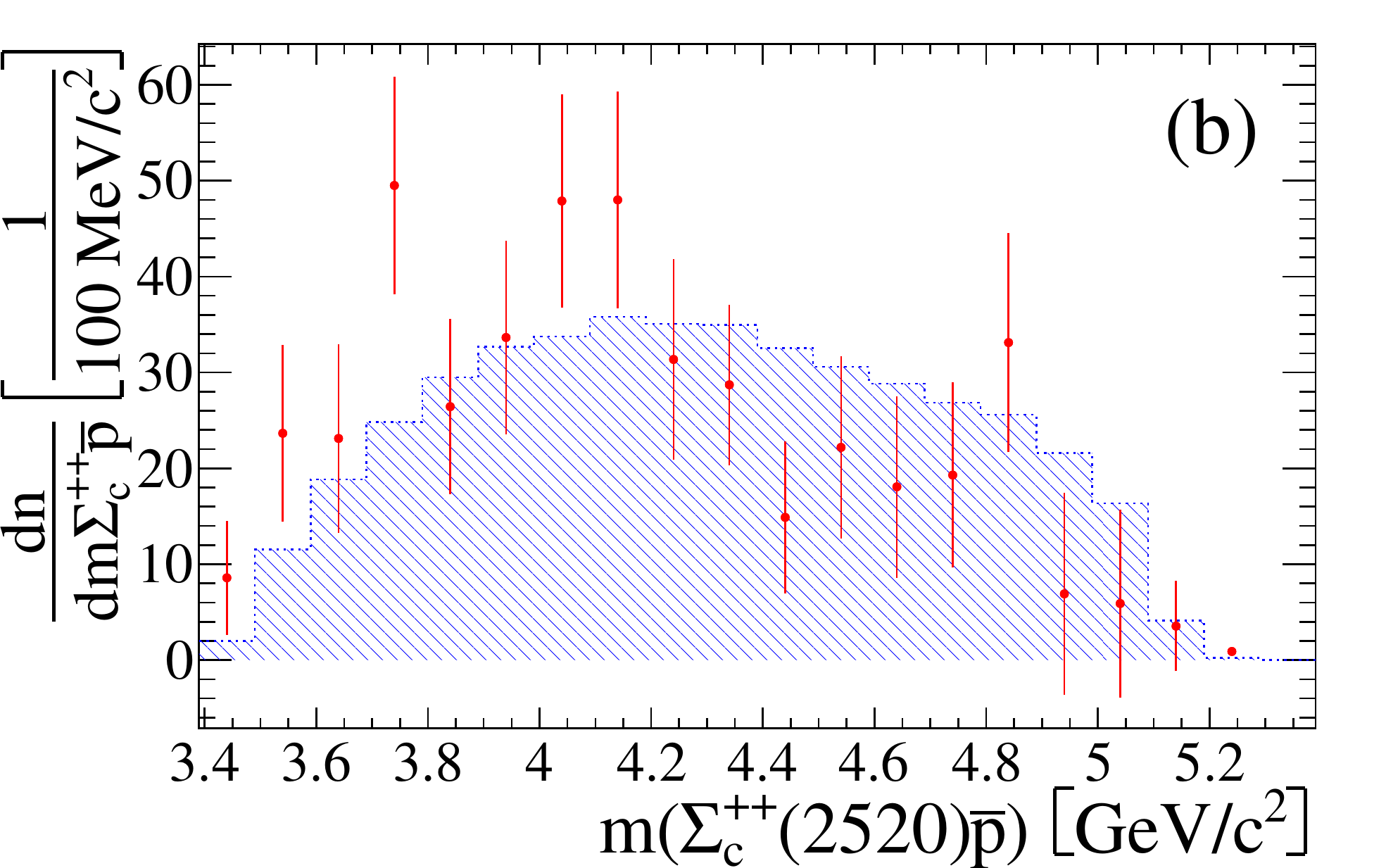}
      \label{fig:SplotsSigmaCplpl2520Vglb}
    }
    \subfigure{
      \includegraphics[width=0.95\textwidth]{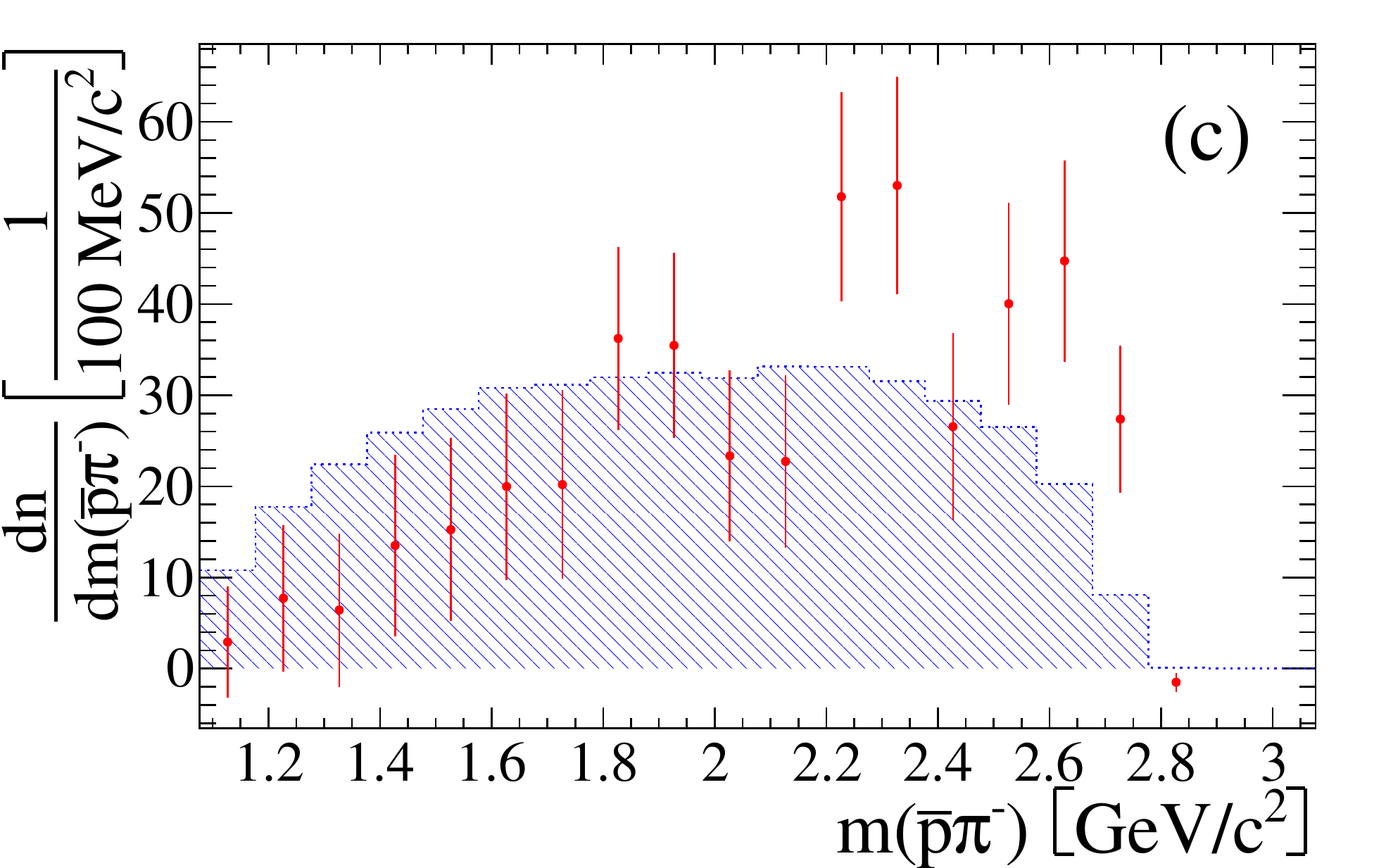}
      \label{fig:SplotsSigmaCplpl2520Vglc}
    }
    \caption{Invariant mass distributions from $\Bzb\to\SigmaC(2520)^{++}\antiproton\pim$ signal events extracted with the \sPlot method.
Data points are displayed in comparison with the distribution of reconstructed phase-space generated $\Bzb\to\SigmaC(2520)^{++}\antiproton\pim$  MC events scaled to the same number of entries (shaded histograms).}
    \label{fig:SplotsSigmaCplpl2520Vgl}
\end{minipage}
\end{figure}

The distributions of signal events are extracted using the \sPlot technique \cite{Pivk:2004ty} in variables other than the discrimination variables: we perform a fit to the two-dimensional distributions of the signal variables \minv and \mpppm where all shape parameters are fixed, and only the signal yields $N_i$ for each signal and background class $i$ are allowed to vary. Distributions for class $i$ are obtained using per-event weights
\begin{equation}
W_i(\minv,\mpppm) = \frac{\sum_{j=1}^{N_{total}} V_{ij}
f_{j}(\minv,\mpppm)}{\sum_{j=1}^{N_{total}} N_{j}f_{j}(\minv,\mpppm)}.
\end{equation}
where $N_{total} = \sum N_i$, the $N_{i}$ and $f_{i}(\minv,\mpppm)$ are the fitted yield and PDF value for the event in the class $i$, and $V_{ij} = \mathop{Cov}(N_i,N_j)$ is the fit's covariance matrix. We use these weights to generate histograms in Dalitz variables of the $\SigmaC \antiproton \pi$ three-body systems. Uncertainties are calculated for \sPlot histograms with $\sqrt{\sum_{i}W_{i}^{2}}$. \\\indent
In Fig.\;\ref{fig:SplotsSigmaC02455Vgla}, $\Bzb\to\SigmaC(2455)^{0}\antiproton\pip$ events are seen to exhibit a sharp enhancement just above the threshold in the $m\left(\SigmaC(2455)^{0}\pip\right)$ distribution; however, there is insufficient information to reliably identify the $\LC(2595)^{+}$ or $\LC(2625)^{+}$ states. In Fig.\;\ref{fig:SplotsSigmaC02455Vglb}, signal events from $\Bzb\to\SigmaC(2455)^{0}\antiproton\pip$  accumulate in $m\left(\SigmaC(2455)^{0}\antiproton\right)$ only for values larger than $3.8\gevcc$, clearly ruling out an enhancement at baryon-antibaryon invariant mass threshold, which has been seen in other decays \cite{Babar_Marcus,Babar_Stephanie,TaeMin_arXiv,BLambdaPPi:2009ru,Aubert:2007uf}. Different behavior is observed in Fig.\;\ref{fig:SplotsSigmaC02455Vglc} where $\Bzb\to\SigmaC(2455)^{0}\antiproton\pip$ events accumulate significantly only for masses $m\left(\antiproton\pip\right)<1.8\gevcc$. Possible sources for this structure could be $\Delta$ baryons or $N^{*}$ nucleon resonances decaying to $\antiproton\pip$. However, due to the overlap of possible broad baryon resonances, we cannot come to a conclusion on specific modes. \\\indent
The corresponding distributions for $\Bzb\to\SigmaC(2455)^{++}\antiproton\pim$ events exhibit different behavior. In the $m\left(\SigmaCplpl\pim\right)$ distribution of Fig.\;\ref{fig:SplotsSigmaCplpl2455Vgla}, no enhancement in the threshold region is visible while a bump at around $2.9\gevcc$ may may be due to additional contributions from intermediate $\LC(2880)^{+}$ and/or $\LC(2940)^{+}$ resonances. In the $m\left(\SigmaC(2455)^{++}\antiproton\right)$ distribution of Fig.\;\ref{fig:SplotsSigmaCplpl2455Vglb}, events with masses below $3.8\gevcc$ contribute prominently, in contrast to the corresponding $m\left(\SigmaC(2455)^{0}\antiproton\right)$ distribution of Fig.\;\ref{fig:SplotsSigmaC02455Vglb}, making  $\Bzb\to\SigmaC(2455)^{++}\antiproton\pim$ decays more similar to other baryonic decays \cite{Babar_Marcus,Babar_Stephanie,TaeMin_arXiv,BLambdaPPi:2009ru,Aubert:2007uf}. In Fig.\;\ref{fig:SplotsSigmaCplpl2455Vglc}, events from $\Bzb\to\SigmaC(2455)^{++}\antiproton\pim$ are distributed in $m\left(\antiproton\pim\right)$ without an obvious structure, unlike events from $\Bzb\to\SigmaC(2455)^{0}\antiproton\pip$ in $m\left(\antiproton\pip\right)$ in Fig.\;\ref{fig:SplotsSigmaC02455Vglc}.\\\indent
The distributions from $\Bzb\to\SigmaC(2520)^{++}\antiproton\pim$ in Fig. \ref{fig:SplotsSigmaCplpl2520Vgl} are similar to the distributions from $\Bzb\to\SigmaC(2455)^{++}\antiproton\pim$ events shifted to higher invariant masses.\\\indent
Due to the relatively small event yield for $\Bzb\to\SigmaC(2520)^{0}\antiproton\pip$ decays, the corresponding  \sPlot{s} are not conclusive, and are therefore not presented.

\section{non-\SigmaC $\Bz\to\LCp\antiproton\pip\pim$ Analysis}\label{txt:nonSigmaC_1} 
The rates of events decaying into the $\Bzb\to\LCp\antiproton\pip\pim$ final state without intermediate $\SigmaC(2455,2520)^{++}$ or $\SigmaC(2455,2520)^{0}$ resonances are determined in one-dimensional fits to \minvn. The non-\SigmaC signal yield components measured in the fits to $\minvn\,{:}\,\mpp$ and $\minvn\,{:}\,\mpm$ are not used, since these yields are correlated.
\subsection{Fit strategy and results}
\begin{figure*}[htpb]
    \centering
     \begin{tabular}{cc}
       \begin{minipage}[H]{0.475\textwidth}
         {
           \includegraphics[width=0.95\textwidth]{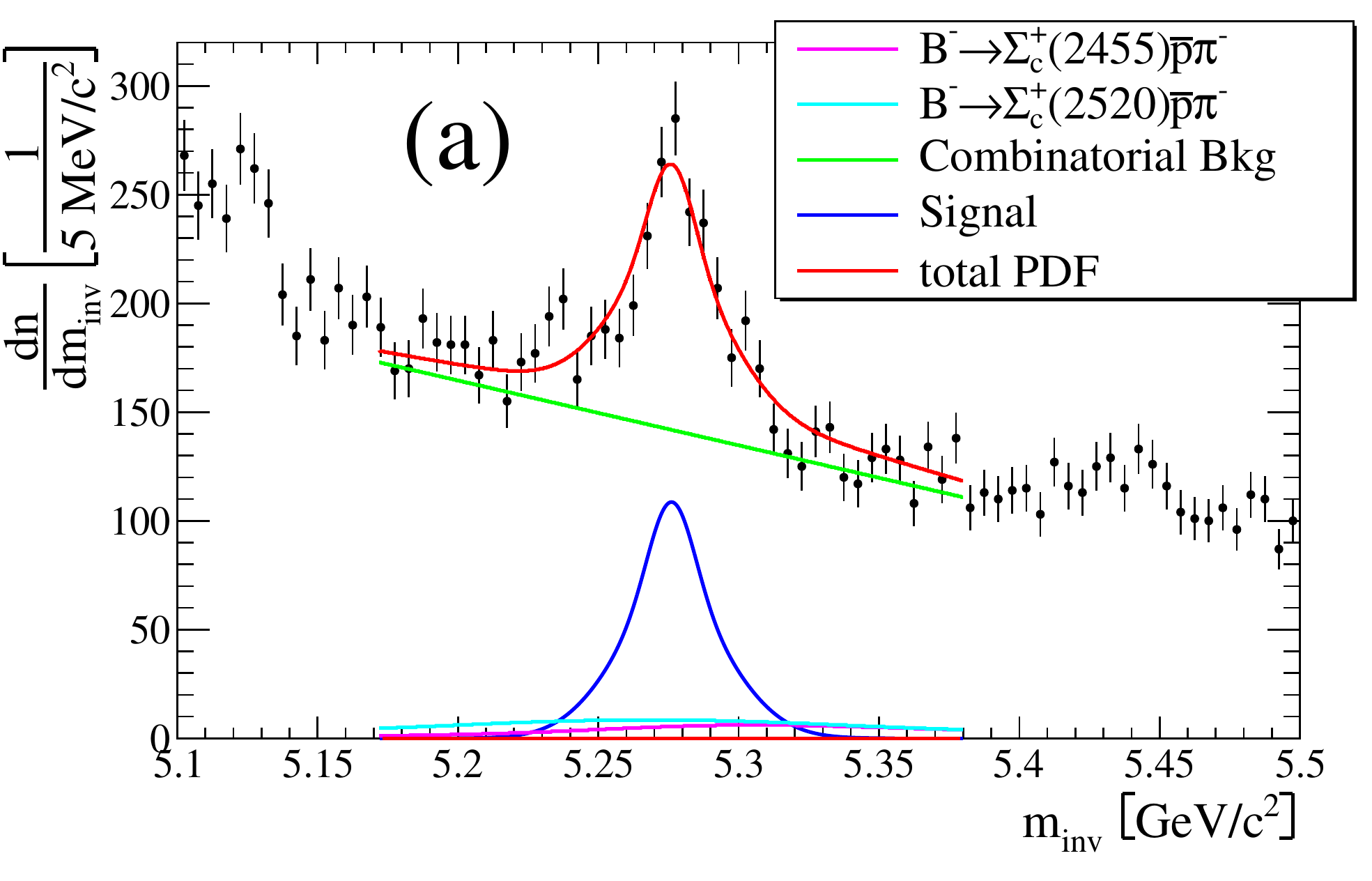}
           \label{fig:FitDatenInvMass_Region1_SigmaCpl_1a}
           }
       \end{minipage}
       &
       \begin{minipage}[H]{0.475\textwidth}
         \includegraphics[width=0.95\textwidth]{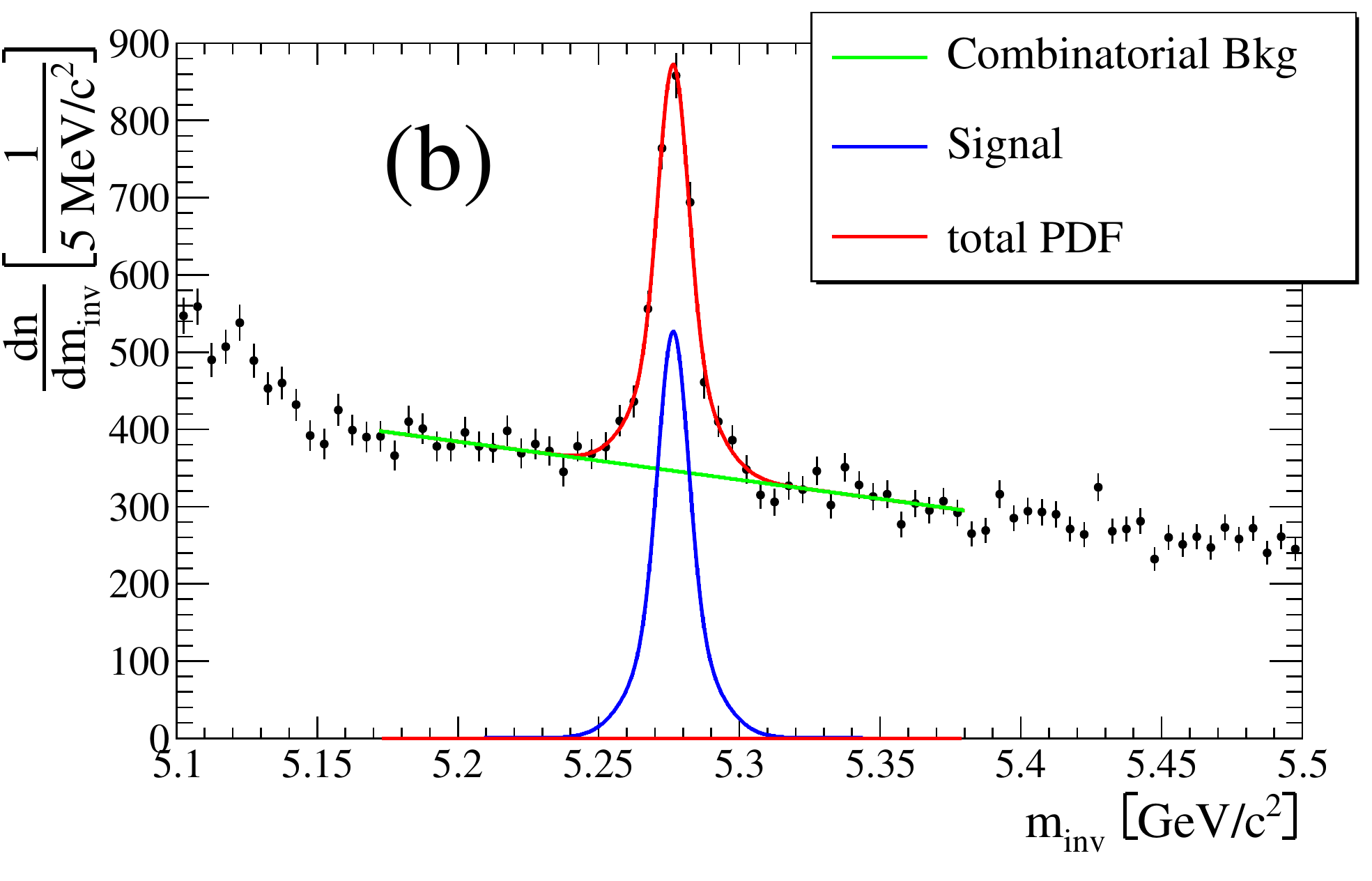}
         \label{fig:FitDatenInvMass_Region1_SigmaCpl_1b}
       
       \end{minipage}
     \end{tabular}
    \caption{Fits to \minv distributions of events decaying into the four-body final state $\Bzb\to\LCp\antiproton\pip\pim$ without intermediate $\SigmaC(2455,2520)^{++,0}$ resonances. In (a) events originate from subset $m_{\mathrm{inv}}^{I}$ excluding the $\SigmaC(2455,2520)^{++,0}$ signal regions in $\m\left(\LCp\pipm\right)$ and taking background from $\Bm\to\SigmaC(2455,2520)^{+}\antiproton\pim$ into account. In (b) the events originate from subset $m_{\mathrm{inv}}^{II}$. }
    \label{fig:FitDatenInvMass_Region1_SigmaCpl_1}
  \end{figure*}

The data are divided into two sets: subset $\m^{I}_{\mathrm{inv}}$ with $\mpp<2.625\gevcc$ and $\mpm<2.625\gevcc$, and subset $m_{\mathrm{inv}}^{II}$ with $\mpp\geq2.625\gevcc$ and $\mpm\geq2.625\gevcc$.  Thus, potential contributions of $\Bm\to\SigmaC(2455,2520)^{+}\antiproton\pim$ are confined to the $m_{\mathrm{inv}}^{I}$ subset. In $m_{\mathrm{inv}}^{I}$, resonant signal events $\Bzb\to\SigmaC(2455,2520)^{++,0}\antiproton\pimp$ are excluded by requiring $m\left(\LCp\pipm\right)$ to lie outside $[2.447,2.461]\gevcc$ and $[2.498,2.538]\gevcc$.\\\indent
In the fit of the distribution of $m_{\mathrm{inv}}^{I}$, background from $\Bm\to\SigmaC(2455)^{+}\antiproton\pim$ decays is taken into account using a double-Gaussian PDF consisting of two single-Gaussian functions with different means and widths. The shape parameters are fixed to values obtained from signal MC, and the signal yield is fixed to the yield measured in the fit to $\minvn\,{:}\,\mpp$ (Table \ref{tab:FitYields_1}). Similarly, background from $\Bm\to\SigmaC(2520)^{+}\antiproton\pim$ events is described by a single-Gaussian PDF with fixed shape parameters from MC and the yield fixed to the fitted yield in $\minvn\,{:}\,\mpp$. Combinatorial background is described with a linear function in \minvn. Signal event contributions are described with a double Gaussian with a shared mean; the parameters are allowed to float. In the distribution of $\m^{II}_{inv}$, the fit is performed with a first-order polynomial for 
background and a double Gaussian with shared mean for signal, since no peaking background is expected here.\\\indent
The fits are shown in Fig.\;\ref{fig:FitDatenInvMass_Region1_SigmaCpl_1} and the yields are given in Table \ref{tab:FitYields_2}.
\begin{table}[htpb]
  \centering
  \caption{Signal yields without efficiency correction for decays without intermediate $\SigmaC(2455,2520)^{++,0}$ resonances. The first uncertainty is the statistical uncertainty. The second uncertainty for $m_{\mathrm{inv}}^{I}$ denotes the uncertainty on the contribution due to ${\Bm\to\SigmaCpl\antiproton\pim}$. \vrule width0pt depth12pt}
  \begin{tabular}{l c }\hline\hline\\[-5pt]
    {Region} & {Signal yield} \\\hline\\[-5pt]
    {{$m_{\mathrm{inv}}^{I}$}}  & $810 \pm 88 \pm 38$\\[5 pt]
    {{$m_{\mathrm{inv}}^{II}$}}  & $1918 \pm 91$\\
    \hline\hline
   \end{tabular}
  \label{tab:FitYields_2}
\end{table}

\subsection{Signal event distributions}
\begin{figure*}[htpb]
  \begin{minipage}[H]{0.95\textwidth}
    \begin{tabular}{l r}
  \begin{minipage}[H]{0.45\textwidth}
    \centering
    \subfigure{
      \includegraphics[width=0.95\textwidth]{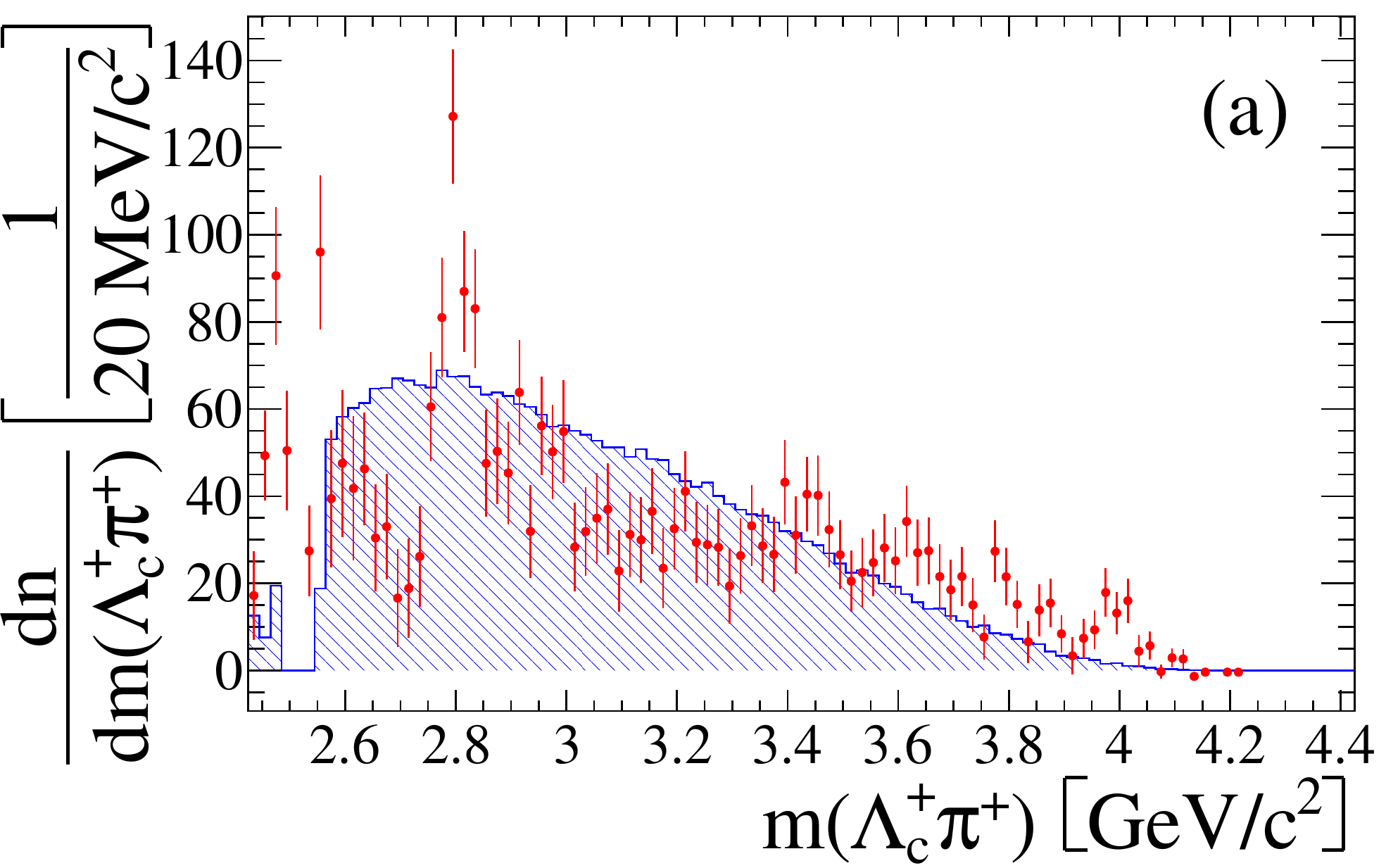}
    \label{SplotsInvMassAusSigmaCFitRegion_1b}
    }
    \subfigure{
      \includegraphics[width=0.95\textwidth]{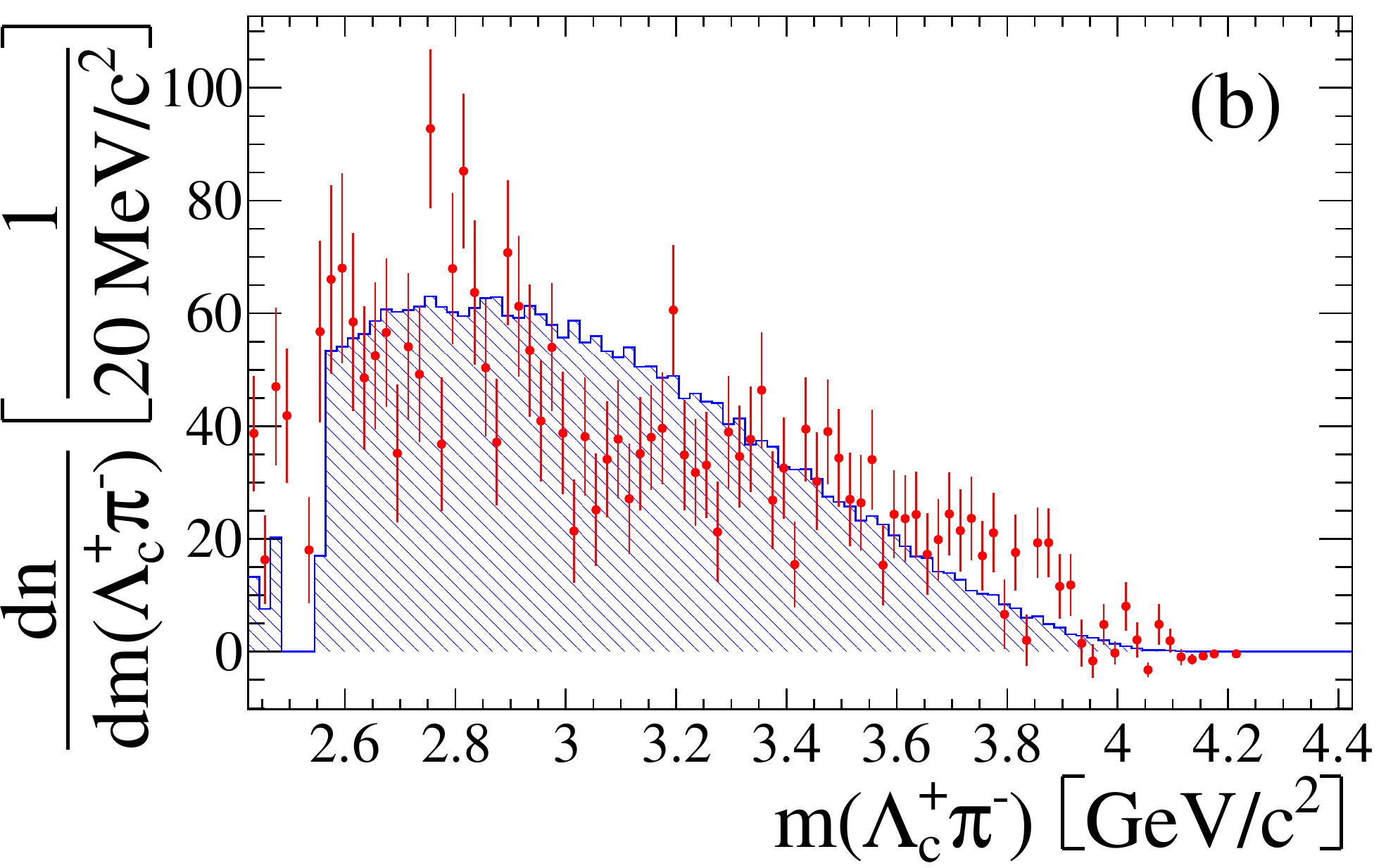}
    \label{SplotsInvMassAusSigmaCFitRegion_1c}
    }
    \subfigure{
      \includegraphics[width=0.95\textwidth]{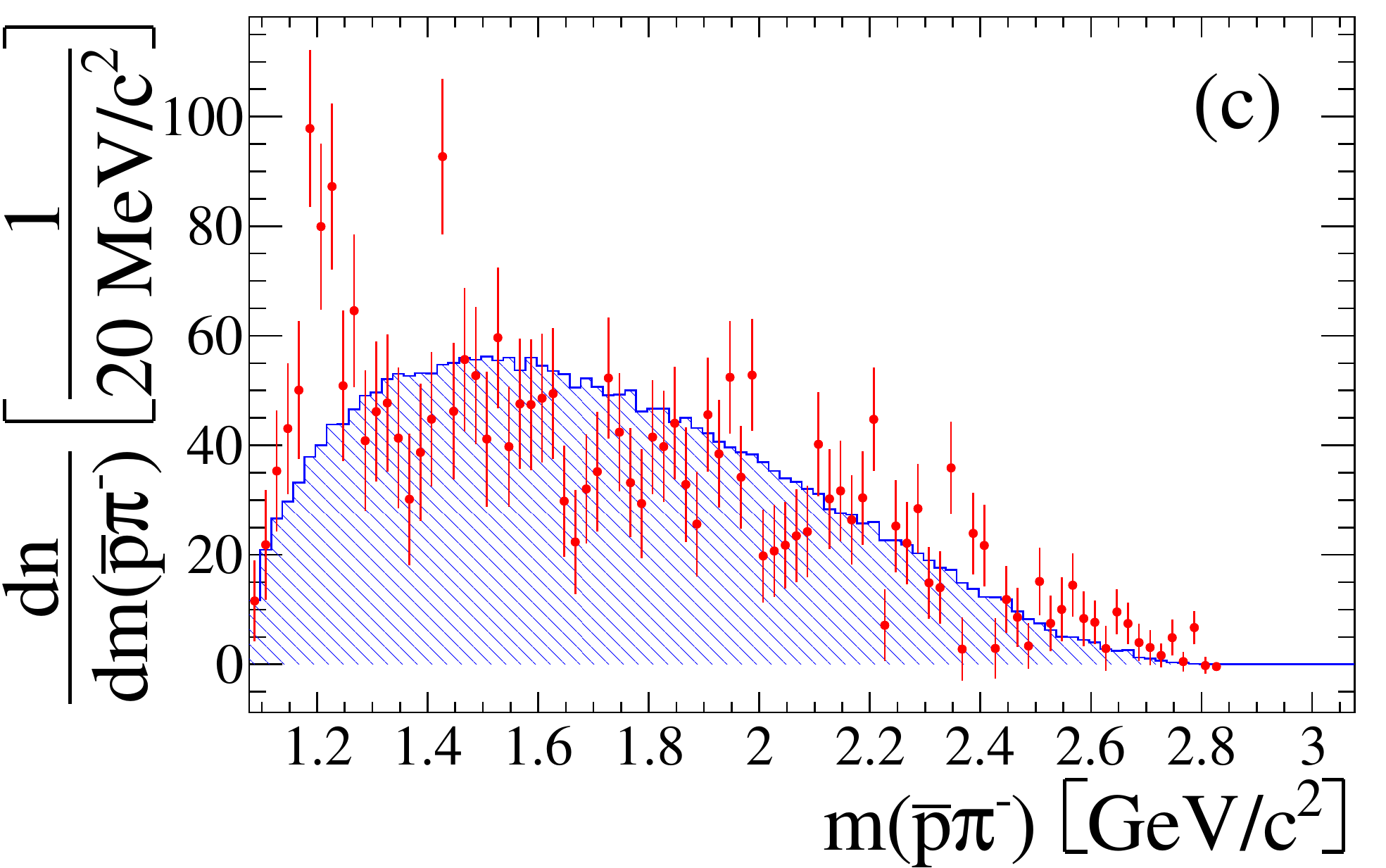}
    \label{SplotsInvMassAusSigmaCFitRegion_1e}
    }
  \end{minipage}
  &
  \begin{minipage}[H]{0.45\textwidth}
    \centering
    \subfigure{
      \includegraphics[width=0.95\textwidth]{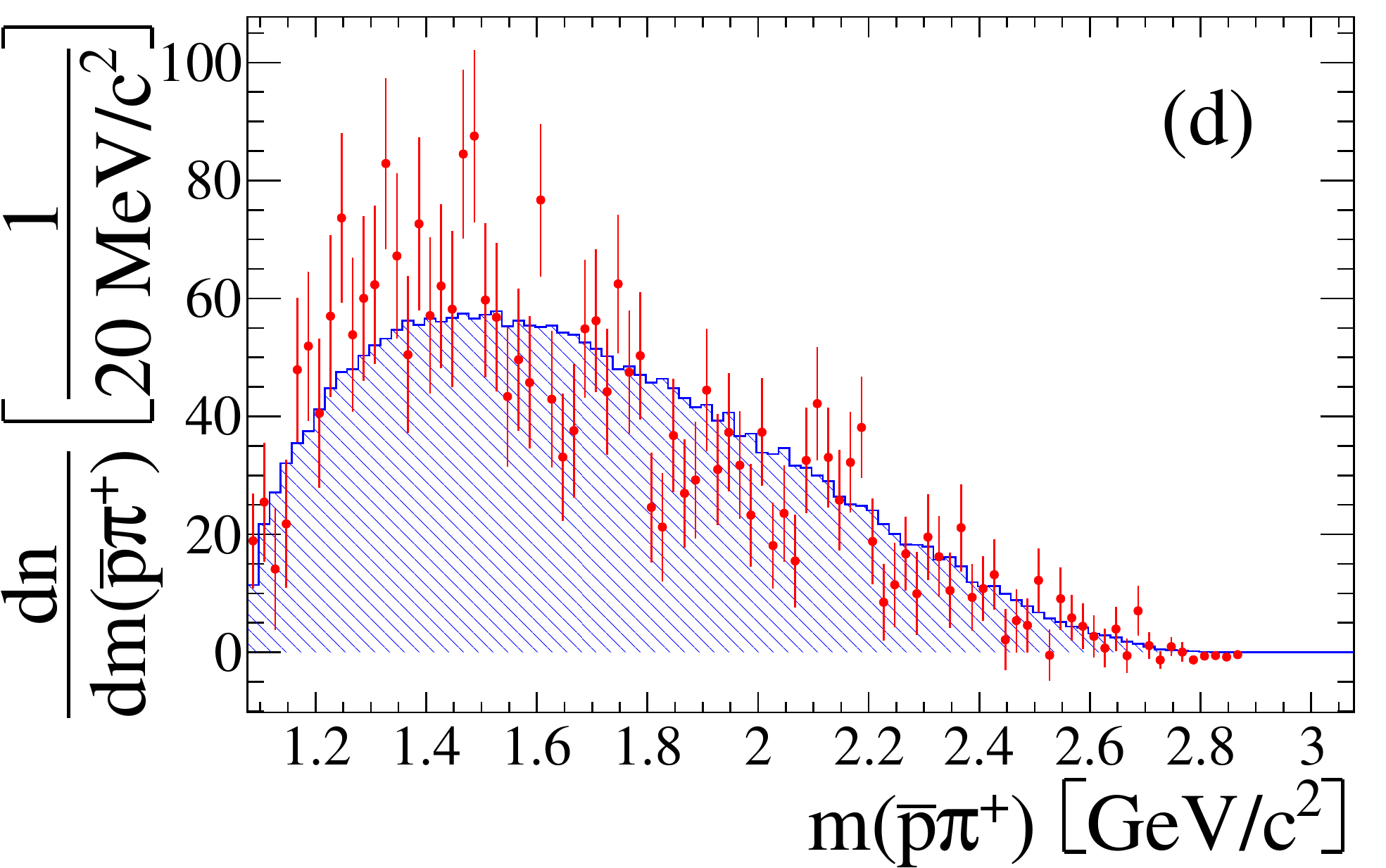}
    \label{SplotsInvMassAusSigmaCFitRegion_1f}
    }
    \subfigure{
      \includegraphics[width=0.95\textwidth]{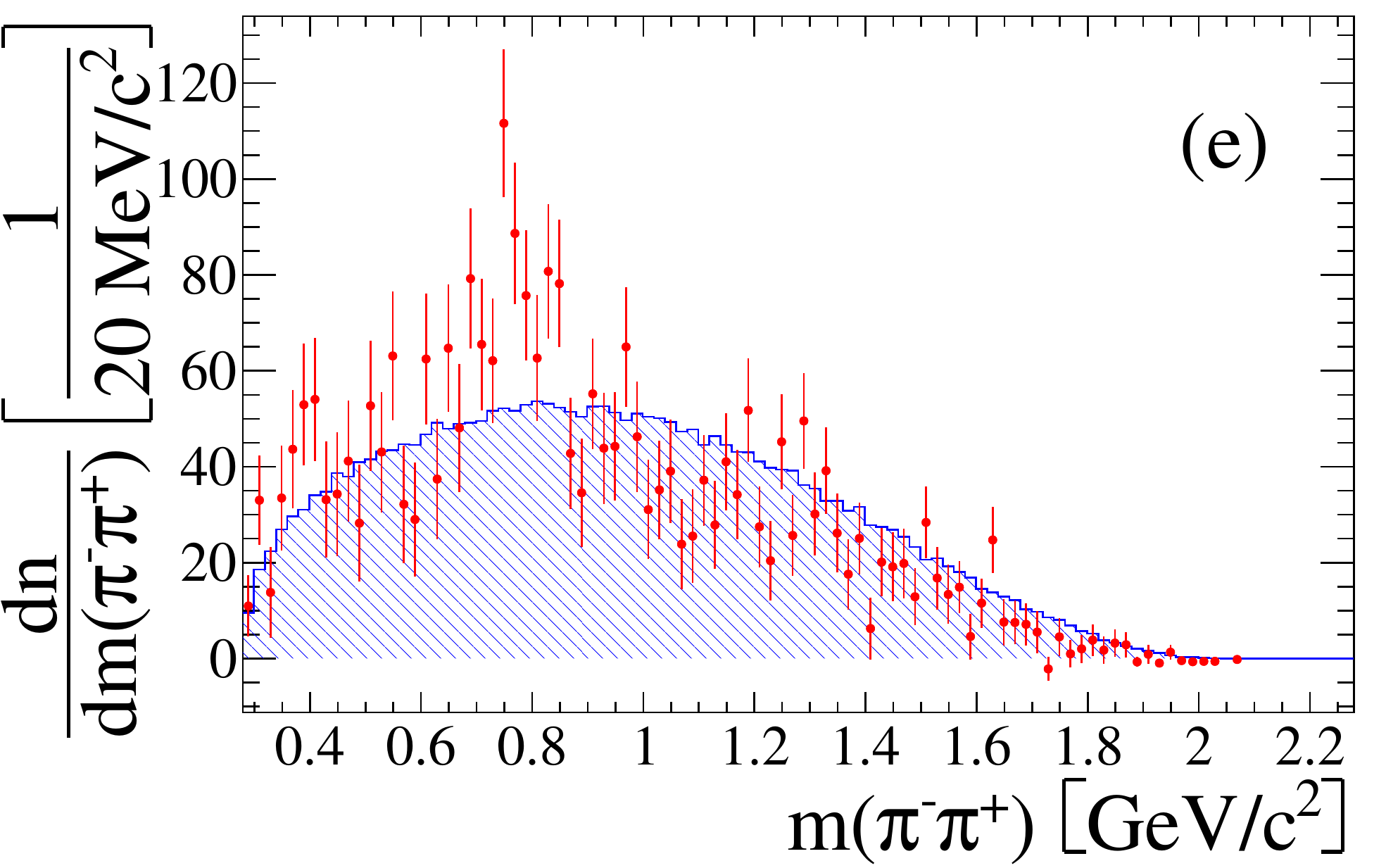}
    \label{SplotsInvMassAusSigmaCFitRegion_1d}
    }
    \subfigure{
      \includegraphics[width=0.95\textwidth]{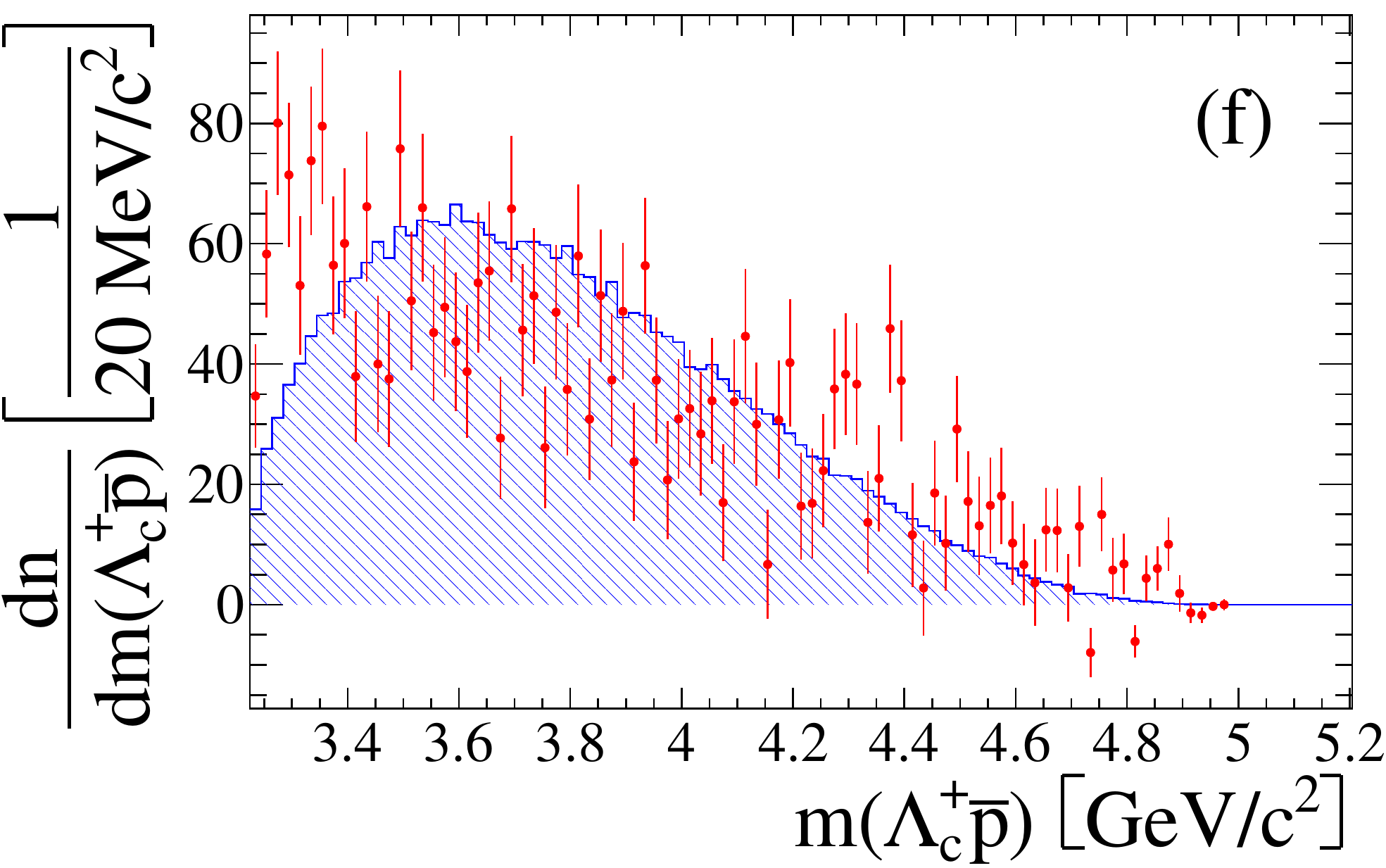}
    \label{SplotsInvMassAusSigmaCFitRegion_1a}
    }
  \end{minipage}
  \end{tabular}
    \caption{Two-body invariant mass distributions for $\Bzb\to\LCp\antiproton\pip\pim$ signal events from the combined subsets $m_{\mathrm{inv}}^{I}$ and $m_{\mathrm{inv}}^{II}$ extracted with the \sPlot method. \SigmaC resonances are vetoed in the respective invariant masses. Data points are displayed in comparison with the distribution of reconstructed phase-space generated $\Bzb\to\LCp\antiproton\pip\pim$  MC events scaled to the same number of entries (shaded histograms).}
    \label{SplotsInvMassAusSigmaCFitRegion_1}
  \end{minipage}
\end{figure*}

\begin{figure}[htpb]
  \begin{minipage}[H]{0.45\textwidth}
    \centering
    \subfigure{
      \includegraphics[width=0.95\textwidth]{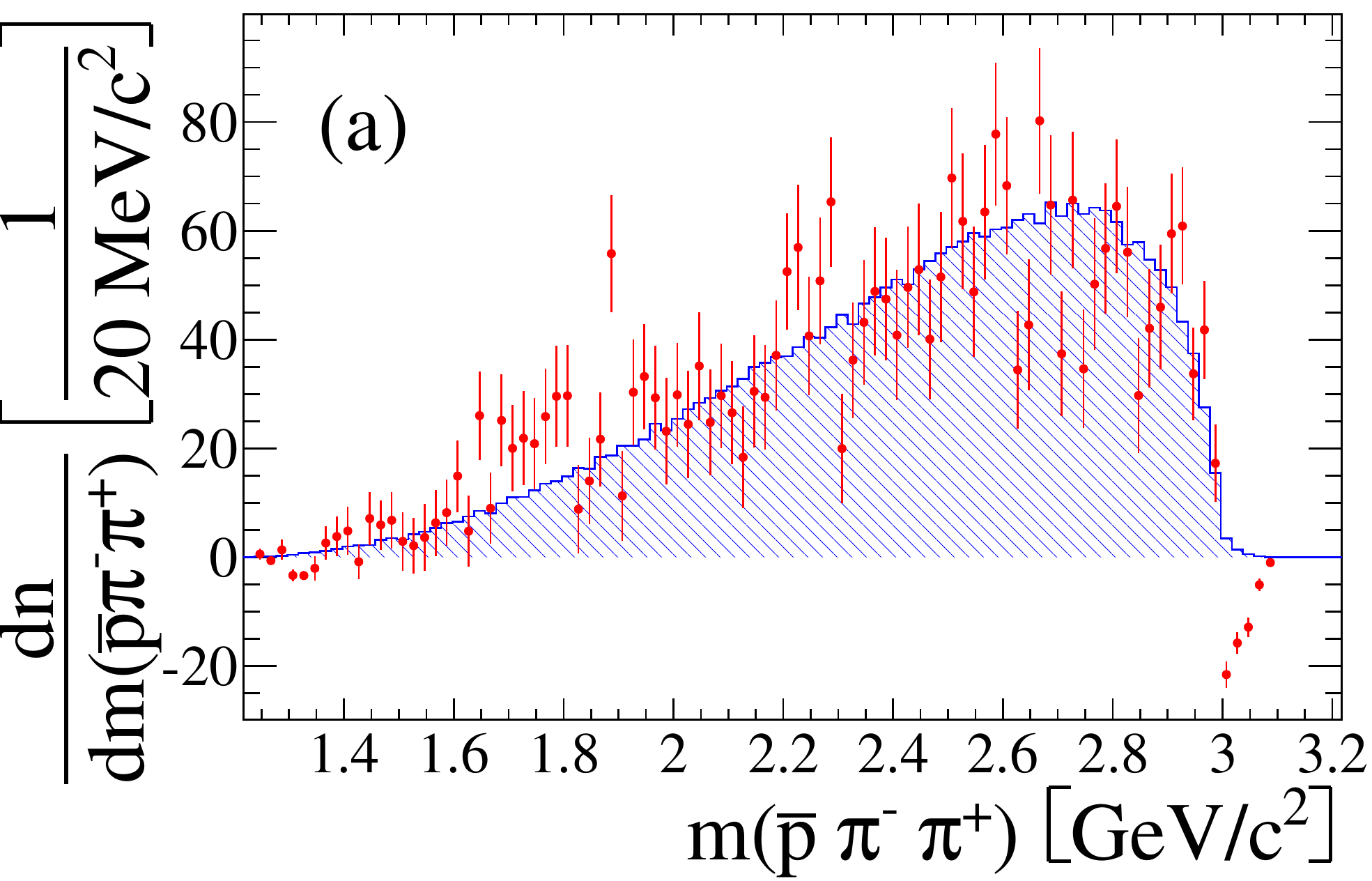}
    \label{SplotsInvMassAusSigmaCFitRegion_2a}
    }
    \subfigure{
      \includegraphics[width=0.95\textwidth]{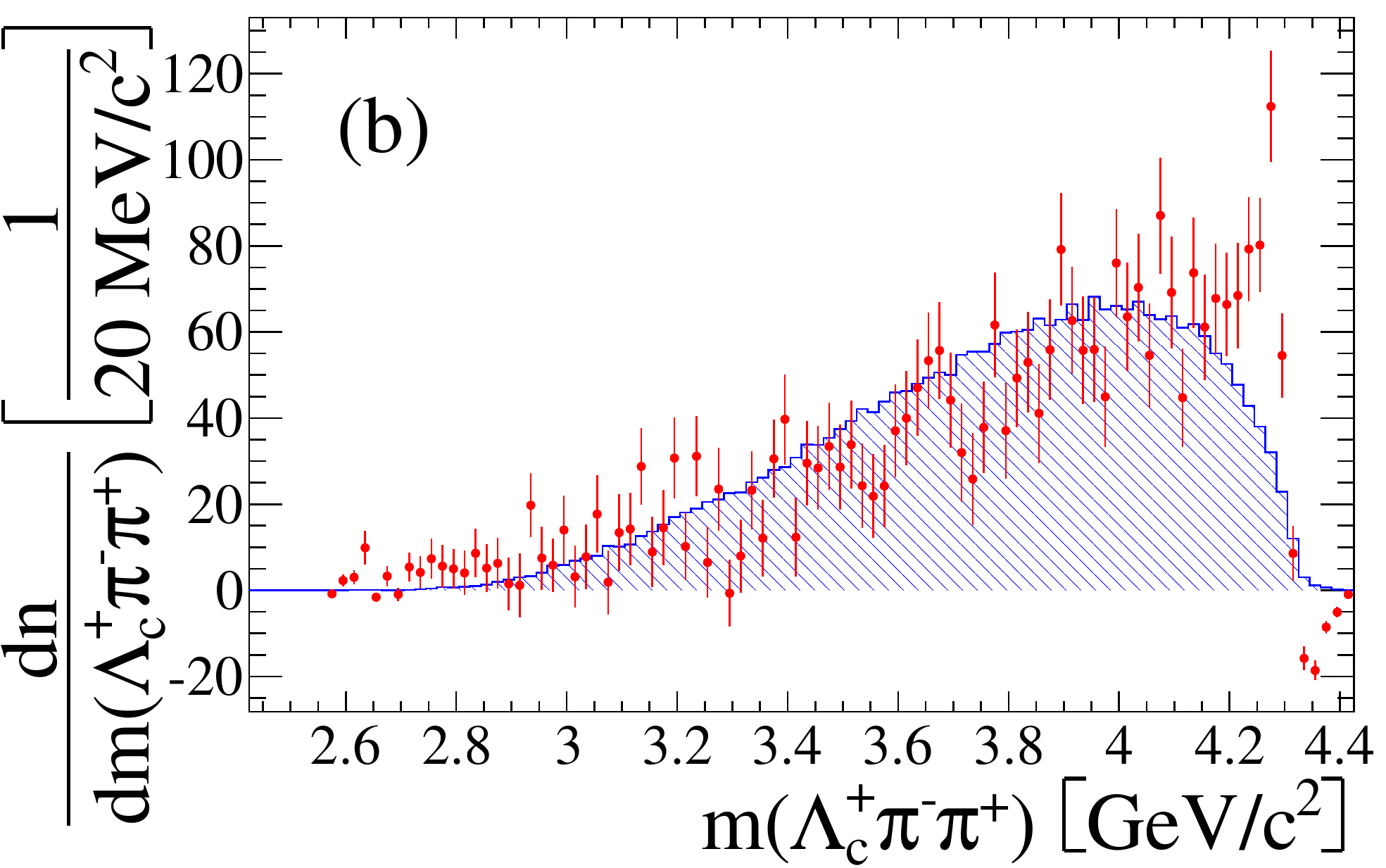}
    \label{SplotsInvMassAusSigmaCFitRegion_2b}
    }
    \subfigure{
      \includegraphics[width=0.95\textwidth]{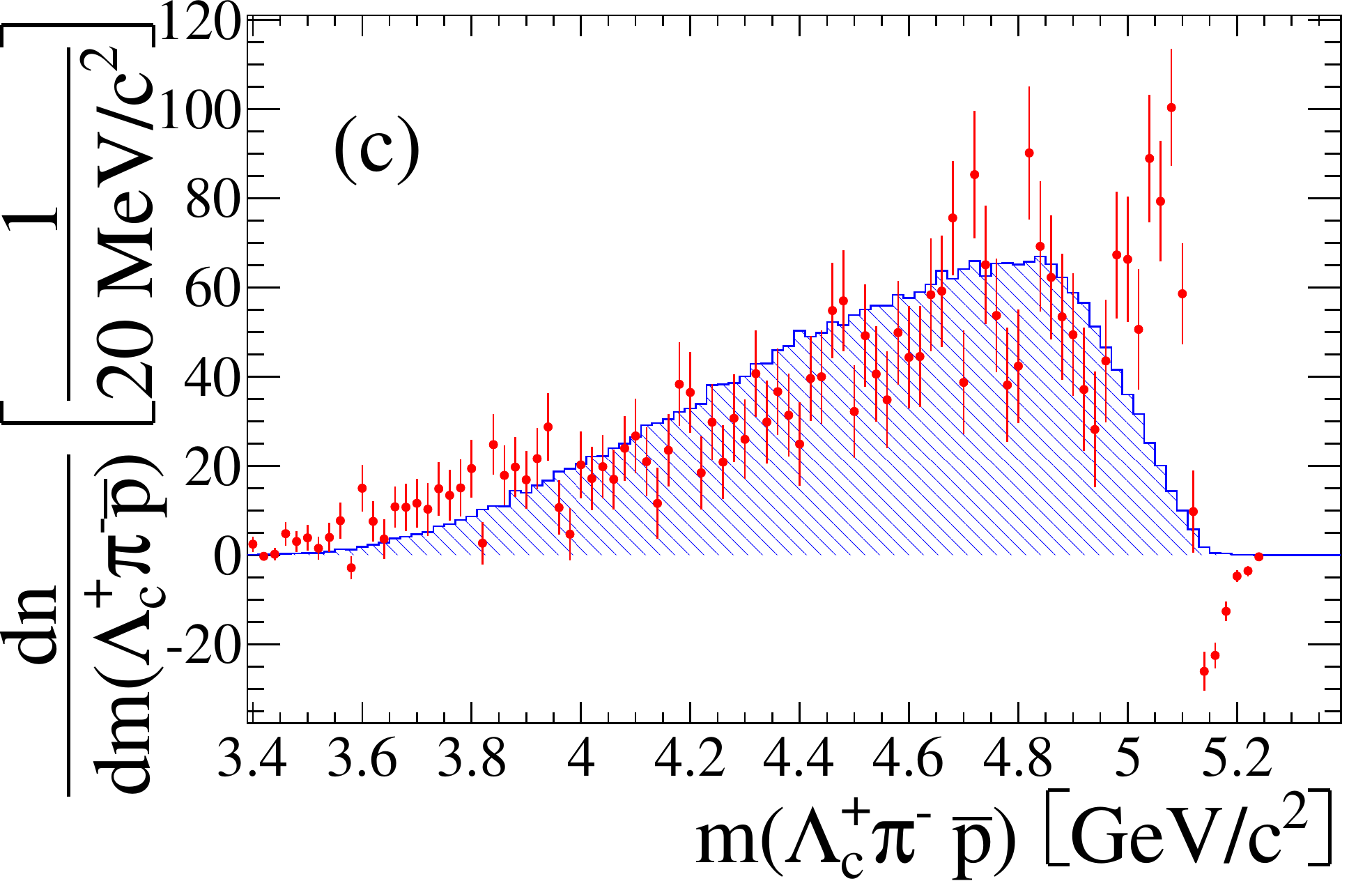}
    \label{SplotsInvMassAusSigmaCFitRegion_2c}
    }
    \subfigure{
      \includegraphics[width=0.95\textwidth]{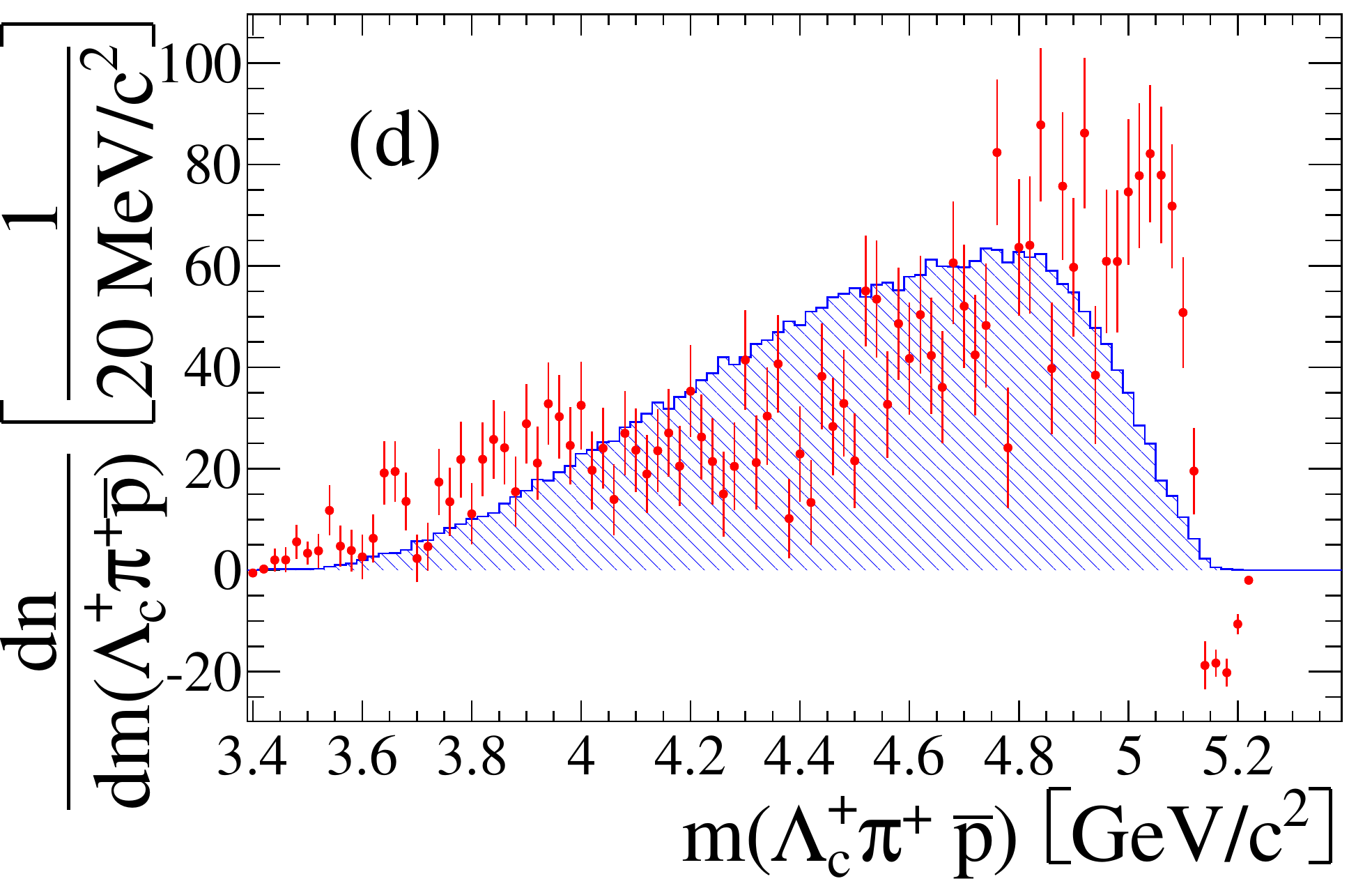}
    \label{SplotsInvMassAusSigmaCFitRegion_2d}
    }
    \caption{Three-body invariant mass distributions for $\Bzb\to\LCp\antiproton\pip\pim$ signal events from the combined subsets $m_{\mathrm{inv}}^{I}$ and $m_{\mathrm{inv}}^{II}$ extracted with the \sPlot method. Data points are displayed in comparison with the distribution of reconstructed phase-space generated $\Bzb\to\LCp\antiproton\pip\pim$  MC events scaled to the same number of entries (shaded histograms).}
    \label{SplotsInvMassAusSigmaCFitRegion_2}
  \end{minipage}
\end{figure}

Figures \ref{SplotsInvMassAusSigmaCFitRegion_1} and \ref{SplotsInvMassAusSigmaCFitRegion_2} show the combined \sPlot{s} for non-\SigmaC $\Bzb\to\LCp\antiproton\pip\pim$ events from the fits to $m_{\mathrm{inv}}^{I}$ and $m_{\mathrm{inv}}^{II}$.\footnote{The fixed background contributions from $\Bm\to\SigmaC(2455,2520)^{+}\antiproton\pim$ are taken into account in the \sPlot weight calculations, following the method described in appendix {\bf{B}} of Ref. \cite{Pivk:2004ty}.} In the $m\left(\LCp\pip\right)$ distribution of Fig.\;\ref{SplotsInvMassAusSigmaCFitRegion_1b}, the contribution from intermediate $\SigmaC(2800)^{++}$ baryons is clearly apparent. The corresponding distribution in $m\left(\LCp\pim\right)$ is shown in Fig.\;\ref{SplotsInvMassAusSigmaCFitRegion_1c}; here the isospin related $\SigmaC(2800)^{0}$ baryon is less significant. Note that the vetoes on low-mass \SigmaC resonances appear as gaps in the distributions. We do not attempt to explicitly measure intermediate states with $\SigmaC(2800)$ baryons with the present approach. 
As described in footnote \ref{footnote_sigmac2800}, significantly differing masses of $\SigmaC(2800)$ resonances have been observed in related $\B\to\LCp\antiproton\pi$ decays \cite{Babar_Marcus,Babar_Stephanie}, which could originate, amongst other possibilities, from different angular momentum states with similar masses or from contamination due to $\B\to D \proton\antiproton \left({{n}} \pi\right)$ decays. Since the present approach uses {\it a priori} information on the masses and widths to generate MC-based PDFs, histogram PDFs cannot be applied for states with uncertain masses or widths.\\\indent
In the distribution of $m\left(\antiproton\pim\right)$ in Fig.\;\ref{SplotsInvMassAusSigmaCFitRegion_1e}, differences are seen compared to the distribution of $m\left(\antiproton\pip\right)$ in Fig.\;\ref{SplotsInvMassAusSigmaCFitRegion_1f}, with events accumulating in  $m\left(\antiproton\pim\right)$ at values near the lower phase-space boundary, suggesting contributions from decays via $\bar{\Delta}^{--}$. Such a structure does not contribute to $m\left(\antiproton\pip\right)$. The $m\left(\pip\pim\right)$ distribution in Fig.~\ref{SplotsInvMassAusSigmaCFitRegion_1d} suggests an intermediate $\rho(770)$ resonance.  However, the data are not sufficiently precise to allow a definite conclusion. 
The $m\left(\LCp\antiproton\right)$ distribution in Fig.\;\ref{SplotsInvMassAusSigmaCFitRegion_1a} shows some enhancement in the baryon-antibaryon mass near threshold, though less strongly than in other measurements with baryonic final states\eg those of Ref. \cite{Babar_Stephanie}. A conclusive interpretation of the $m\left(\LCp\antiproton\right)$ result is difficult, because the MC distribution uses all events in the allowed phase space to avoid a possible bias, averaging over all possible structures. Furthermore, the projections onto the axes of the Dalitz space for the four-final-state-particle system make it difficult to identify reflections from resonances in other invariant masses.\\\indent
The three-body mass distributions are shown in Fig.\;\ref{SplotsInvMassAusSigmaCFitRegion_2}. Here, we do not observe structures in lower invariant-mass ranges that could hint at resonances\eg excited \LC baryons in $m\left(\LCp\pip\pim\right)$.
The bins near the edges of the distribution are not reproduced correctly by the \sPlot method, and show artificial undershoots. This is because the \sPlot technique relies on target variables that are uncorrelated with the discriminating variables. This does not hold for points near the edges of phase space, where there is a dependence on \minv.

\section{Efficiency}\label{txt:Efficiency_1} 
The efficiency of the reconstruction is determined separately for each resonant signal mode and for the non-\SigmaC signal decays. Since signal MC events are generated uniformly in phase space, the observed decay dynamics are not reproduced. To avoid bias from phase-space-dependent reconstruction efficiencies, the MC samples are iteratively reweighted according to the \sPlot histograms ${{\mathcal{N}}}\left[m_{a\,b}\right]_{\mathrm{data}}$ of invariant masses for each signal class from the \B  meson daughters $a$, $b$,$\hdots$. The reweighting is performed iteratively over all two-daughter combinations for each three-body final state and over the three-body combinations for the four-body non-\SigmaC final states. Since we do not observe any non-trivial structures in the smooth and moderately varying efficiency distribution before reweighting, we assume that the projections onto the overdetermined Dalitz variables are sufficient for reweighting.\\\indent
In the initial step $i=1$, the weight is calculated from the \sPlot histogram and the signal MC histogram ${\mathcal{N}}\left[m_{a\,b}\right]_{\mathrm{MC}}^{i=1}$ as  $w^{i=1}\left[m_{a\,b}\right] = {\mathcal{N}\left[m_{a\,b}\right]_{\mathrm{data}}}\,/\,{{\mathcal{N}}\left[m_{a\,b}\right]_{\mathrm{MC}}^{n=1}}$ and applied to each signal MC event. In iteration step $n$, weights are calculated accordingly from the reweighted signal MC from iteration $n-1$ for invariant mass ${m_{b\,c}}$ with $w_{b\,c}={{\mathcal{N}}\left[m_{b\,c}\right]_{\mathrm{data}}}\,/\,{{\mathcal{N}}\left[m_{b\,c}\right]_{\mathrm{MC}}^{n-1}}$. In the following, the signal MC events are weighted by $w_{b\,c}$. Since the MC event values processed in step $n$ originate from weighting the MC events in the previous step $n-1$, the effective weight is $w^{n}=w^{n-1}\,\cdot\,w_{b\,c}$. Negative \sPlot  entries are set to zero to avoid nonphysical weightings.\\\indent
After each step, a \chiq fit of the \minv distribution of the reweighted signal MC is performed to obtain the number of reconstructed weighted events. Thus, the reconstruction efficiency is calculated from the number of weighted reconstructed events and the sum of weighted generated events.  When the reconstruction efficiencies in the last steps of one cycle through all \B  daughter combinations are compatible with each other within the uncertainties, we assume that the reconstruction efficiency has converged and stop the iteration. The reconstruction efficiencies are listed in Table \ref{tab:ReconstructionEfficiencies_1}. The differences between unweighted and weighted efficiencies vary between $2.1\%$ for $\Bzb\to\SigmaC(2455)^{++}\antiproton\pim$ and $7.9\%$ for $\Bzb\to\SigmaC(2455)^{0}\antiproton\pip$.
\begin{table}[htpb]
  \centering
  \caption{Reconstruction efficiencies for each signal event class, after reweighting signal MC events according to \sPlot  distributions.\vrule width0pt depth12pt}
  \begin{tabular}{l c}\hline\hline
    {Mode/region} & {Reconstruction efficiency} \\\hline 
     $\Bzb\to\SigmaC(2455)^{0}\antiproton\pip$ & $\left(16.4 \pm 0.3\right)\%$ \\
    $\Bzb\to\SigmaC(2520)^{0}\antiproton\pip$ & $\left(16.8 \pm 0.3\right)\%$ \\
    $\Bzb\to\SigmaC(2455)^{++}\antiproton\pim$ & $\left(14.5 \pm 0.1\right)\%$ \\
    $\Bzb\to\SigmaC(2520)^{++}\antiproton\pim$ & $\left(17.0 \pm 0.2\right)\%$ \\
    $\Bzb\to\LCp\antiproton\pip\pim_{m_{\mathrm{inv}}^{I}}$ & $\left(11.6\pm 0.7\right)\%$ \\
    $\Bzb\to\LCp\antiproton\pip\pim_{m_{\mathrm{inv}}^{II}}$ & $\left(16.9 \pm 0.1\right)\%$ \\
\hline\hline
  \end{tabular}
  \label{tab:ReconstructionEfficiencies_1}
\end{table}
\section{Branching Fractions}\label{txt:BranchingRatios_1} 
The product branching fractions $\BR$ are calculated for each signal mode $i$ with
\begin{equation}\BR\left[\Bzb\to{\textcolor{grey}{\big[}}\LCp\pipm{\textcolor{grey}{\big]_{\SigmaC}}}\antiproton\pimp\right]_{i}\cdot\BR\left[\LCp\to\proton\Km\pip\right] = \frac{N_{\mathrm{i}}}{N_{\BBbar}} \cdot\frac{1}{\varepsilon_{i}}\end{equation}
 where $N_{\mathrm{i}}$ is the sum of signal-event numbers (Tables \ref{tab:FitYields_1} and \ref{tab:FitYields_2}), $\varepsilon_{i}$ the reconstruction efficiency (Table \ref{tab:ReconstructionEfficiencies_1}), and $N_{\BBbar}$ the number $N\left(\Y4S\right)$ of $\Y4S$ decays; we assume that $\frac{N\left(\Y4S\to\BzBzb\right)}{N\left(\Y4S\right)}=0.50$. For an integrated luminosity of ${\mathcal{L}} = 426\invfb$, $N_{\BBbar} = \left(467.36 \pm  0.11\comment{_{\mathrm{stat}}} \pm 5.14\comment{_{\mathrm{sys}}}\right)\times10^{6}$, where the uncertainties are statistical and systematic. \SigmaC resonances are assumed to decay exclusively into a $\LCp\pi$ pair: $\BR\left[\SigmaC(2455,2520)^{++,0}\to\LCp\pipm\right]\approx 100\%$ \cite{PDG2010}. After accounting for the reconstruction efficiencies, the number of events from the two non-$\SigmaC(2455,2520)$ decay measurements are summed and their statistical uncertainties are added in quadrature.
\section{Systematic Uncertainties}\label{txt:systematics_1}
Systematic uncertainties applying to all modes, such as the uncertainty on ${N_{\BBbar}}$, as well as systematic uncertainties specific only to certain modes, are considered. Table \ref{tab:SysSummary_1} lists the relative systematic uncertainties $u_{x}=\frac{\delta N_{x}}{N_{x}}$ for each uncertainty $x$. Systematic uncertainties on the reconstruction efficiency of the six charged final-state tracks are added linearly to obtain a total tracking uncertainty. One of the largest systematic uncertainties originates from the particle identification efficiencies. The uncertainties are evaluated using MC events, with corrections derived from control samples in the data. In addition, MC events are examined without corrections. The relative difference in the particle identification efficiencies with and without the corrections defines the uncertainty. As discussed above, $\Bzb\to D \proton\antiproton \left({{n}}\pi\right)$ decays and decays through charmonia states $\Bzb\to\left(\c\cbar\right)\Kstarzb\left[\pip\pim\right]$ can yield the same combination of final-state particles as signal events. Based on the known branching fractions \cite{PDG2010}, a total of at most 4.5 events from these two event classes are expected to satisfy the signal selection criteria. Here, a conservative reconstruction efficiency of $\varepsilon = 0.1\%$ is assumed, overestimating the measured efficiencies in signal MC.  The corresponding systematic uncertainty $\left({\B\to D+X,\, \ccbar+X}\right)$ is set equal to 100\% of the corresponding estimated background in line 4 of Table \ref{tab:PseudoMinvBG_1}, for each mode separately.\\\indent 
Only the fits of decays via \SigmaC resonances are affected by an uncertainty on the shape of non-\SigmaC $\Bz\to\LCp\antiproton\pip\pim$ events in the $\minvn\,{:}\,\mpppm$ planes. The parameters on the signal width are varied individually by one standard deviation and the maximum deviation in the event yield is taken as the systematic uncertainty (labeled ``${\mathrm{nonres.\,shape}}$'' in Table \ref{tab:SysSummary_1}). For the shape of combinatorial background, a systematic uncertainty is determined by varying in the PDF the constant describing the end point of the phase space, $e_{\mathrm{up}}$. Fits are repeated with the constant moved from the nominal upper phase-space limit towards the upper end of the fit region  in 0.2\gevcc steps. The maximal deviations in the fitted signal event yields are taken as the systematic uncertainty, labeled ``${\mathrm{Combi\,Bkg\,shape}}$''.
For histogram PDFs, systematic uncertainties are negligible due to the large size of the input MC data sets. In studies on control variables, we found a good agreement between data and MC-generated events. For deviations of the MC-generated events from data, we applied corrections as described in \ref{txt:HistoPDFVerification}. An uncertainty, labeled ``${\mathrm{Eff.\;Corr.}}$'', on the efficiency-calculation weighting is evaluated after completing a cycle through all daughter combinations for each mode. The values converge and are within the statistical uncertainties for all modes after one full cycle, except for $\Bzb\to\SigmaC(2520)^{0}\antiproton\pip$. For this mode, the efficiencies differ by 1.9\% with a statistical uncertainty of 1.8\% and the difference is taken as additional systematic uncertainty.\\\indent
A systematic uncertainty, labeled ``${\Bm\,(I+II)}$'', on the contribution of $\Bm\to\SigmaC(2455,2520)^{+}\antiproton\pim$ events to the non-\SigmaC $\Bzb\to\LCp\antiproton\pip\pim$ yields is calculated by repeating fits to the $m_{\mathrm{inv}}^{I}$ distribution assuming no contribution and overestimating the found contribution by a factor of two. The maximum deviation of $38$ events is included as systematic uncertainty.
\begin{table*}[htpb]
\centering
\caption{Summary of relative systematic uncertainties $u_{i}=\frac{\delta N_{i}}{N_{i}}$ in [\%] for {non-\SigmaC} $\Bzb\to\LCp\antiproton\pip\pim$, resonant $\Bzb\to\SigmaC(2455)\antiproton\pipm$, and resonant $\Bzb\to\SigmaC(2520)\antiproton\pipm$ decays. The total systematic uncertainties, added in quadrature, are given in the last row.\vrule width0pt depth12pt}
\label{tab:SysSummary_1}
\begin{tabular}{l c c c c c  }\hline\hline
 {Uncertainties $u_{i}$ [\%]}                            & {non-\SigmaC}  & { $\SigmaC(2455)^{0}$} & { $\SigmaC(2455)^{++}$}  & { $\SigmaC(2520)^{0}$} & { $\SigmaC(2520)^{++}$} \\\hline
${N_{\BBbar}}$                    & 1.1              & 1.1                 & 1.1                  & 1.1                 & 1.1                 \\
${\mathrm{Tracking}}$                                & 1.2             & 1.2                & 1.2                 & 1.2                & 1.2                \\
${\mathrm{PID}}$                                     & 4.3              & 4.3                 & 4.3                  & 4.3                 & 4.3                 \\
${\B\to D+X,\, \ccbar+X}\quad$ &  0.1            & 0.5                 & 0.5                  & 0.5                 & 0.5                 \\
${\mathrm{nonres.\,shape}}$                 &                    & 0.1                 & 0.1                  & 0.4                 & 0.4                 \\
${\mathrm{Combi\,Bkg\,shape}}$              &                   & 0.001             & 0.001              & 0.7                & 0.7                \\
${\mathrm{Eff.\;Corr.}}$                    &                    &                       &                        & 0.1                 &                      \\
${\Bm\,(I+II)}$                             & 1.8              &                       &                        &                       &                      \\[10 pt]
$\sqrt{\sum{u_{i}^{2}}}$                      & 4.9             & 4.6                & 4.6                 & 4.7                & 4.7                \\[3 pt]
\hline\hline
\end{tabular}
\end{table*}
\section{Results}\label{txt:Results_1}
Table \ref{tab:BranchingFractionsInclSystematics_1} lists the measured branching fractions for each mode. Due to the large uncertainty on $\BR\left[\LCp\to\proton\Km\pip\right]$, its systematic uncertainty is given separately. The uncertainties on the total branching fraction of all $\Bzb\to\LCp\antiproton\pip\pim_{\mathrm{total}}$ decays are added quadratically when uncorrelated and linearly when correlated.\\\indent
Because the $\Bzb\to\SigmaC(2520)^{0}\antiproton\pip$ signal has less than three standard deviations significance, we also report a 90\% confidence level (CL) upper limit for this  channel.  The upper limit is determined using Bayesian methods, with statistical and systematic uncertainties added in quadrature. We do not include the uncertainty on $\BR\left[\LCp\to\proton\Km\pip\right]$ in the systematic uncertainty of the upper limit, but factor out the current branching fraction of $0.05$ \cite{PDG2010}. Assuming a Gaussian distribution, the 90\% integral of the physically meaningful region $\BR\geq0$ yields
\begin{eqnarray}
\BR[\Bzb\to\SigmaC(2520)^{0}\antiproton\pip]\cdot\frac{\BR\left[\LCp\to\proton\Km\pip\right]}{0.05}\\\nonumber 
< 3.10 \times 10^{-5}
\label{math:UpperLimitSigmaCz2520}
\end{eqnarray}
at $90\%$ confidence level.\\\indent
Resonant decays via $\SigmaC(2455,2520)$ baryons provide a large contribution to the four-body final state. The ratios of decays via \SigmaC resonances in comparison to the largest such mode, $\Bzb\to\SigmaC(2455)^{++}\antiproton\pim$, are
\begin{eqnarray}
\frac{\BR\left[\Bzb\to\SigmaC(2455)^{0}\antiproton\pip\right]}{\BR\left[\Bzb\to\SigmaC(2455)^{++}\antiproton\pim\right]} = 0.425 \pm 0.036,\\
\frac{\BR\left[\Bzb\to\SigmaC(2520)^{++}\antiproton\pim\right]}{\BR\left[\Bzb\to\SigmaC(2455)^{++}\antiproton\pim\right]} = 0.541 \pm 0.052,
\end{eqnarray}
while the fraction of all decays that proceed via the $\SigmaC(2455)^{+}\antiproton\pim$ mode is
\begin{eqnarray}
\frac{\BR\left[\Bzb\to\SigmaC(2455)^{++}\antiproton\pim\right]}{\BR\left[\Bzb\to\LCp\antiproton\pip\pim\right]_{\mathsf{total}}}= 0.174 \pm 0.047.
\end{eqnarray}
In these three results, systematic uncertainties common to numerator and denominator cancel, and only the systematic uncertainties specific to each mode are added in quadrature. The three-body intermediate states have comparable branching fractions to the non-resonant three body decays $\B\to\LCp\antiproton\pi$ \cite{Babar_Marcus,Babar_Stephanie,BELLE_SigmaC,Lit:Dytman:2002yd,Babar_Oliver}.\\\indent
The measured branching fractions are in good agreement with previous measurements from Belle \cite{BELLE_SigmaC}.
\begin{table*}[htpb]
  \centering
\caption{Branching fractions of the resonant decays $\Bzb\to\SigmaC(2455,2520)^{++,0}\antiproton\pimp$ and  non-$\SigmaC(2455,2520)$ decays $\Bzb\to\LCp\antiproton\pip\pim$ where the first uncertainty is statistical, the second systematic and the third due to the uncertainty on the $\BR\left[\LCp\to\proton\Km\pip\right]$ branching fraction.\vrule width0pt depth12pt}
\begin{tabular}{l c c}\hline\hline
  { Mode/region}                             &  { $\BR\left[\Bzb\right]\cdot\BR\left[\LCp\to\proton\Km\pip\right]\;\left[10^{-6}\right]$}   &  { $\BR\left[\Bzb\right]\;\left[10^{-5}\right]$}   \\\hline
$\Bzb\to\SigmaC(2455)^{0}\antiproton\pip$            & $\left(4.5 \pm 0.3\comment{_{\mathrm{stat}}}   \pm 0.2\comment{_{\mathrm{sys}}}  \right)$                     & $\left(9.1 \pm 0.7\comment{_{\mathrm{stat}}}  \pm 0.4\comment{_{\mathrm{sys}}}   \pm 2.4\comment{_{\LCp}}\right)$\\
$\Bzb\to\SigmaC(2455)^{++}\antiproton\pim$         & $\left(10.7 \pm 0.5\comment{_{\mathrm{stat}}} \pm 0.5\comment{_{\mathrm{sys}}} \right)$                     & $\left(21.3 \pm 1.0\comment{_{\mathrm{stat}}} \pm 1.0\comment{_{\mathrm{sys}}}   \pm 5.5\comment{_{\LCp}}\right)$\\
$\Bzb\to\SigmaC(2520)^{0}\antiproton\pip$            & {{$\left(1.1 \pm 0.4\comment{_{\mathrm{stat}}}  \pm 0.1\comment{_{\mathrm{sys}}}   \right)$}} & {{$\left(2.2 \pm 0.7\comment{_{\mathrm{stat}}}  \pm 0.1\comment{_{\mathrm{sys}}}   \pm 0.6\comment{_{\LCp}}\right)$}}\\
$\Bzb\to\SigmaC(2520)^{++}\antiproton\pim$         & $\left(5.8 \pm 0.5\comment{_{\mathrm{stat}}}   \pm 0.3\comment{_{\mathrm{sys}}}  \right)$                     & $\left(11.5\pm 1.0\comment{_{\mathrm{stat}}} \pm 0.5\comment{_{\mathrm{sys}}}    \pm 3.0\comment{_{\LCp}}\right)$\\[5 pt]
$\Bzb\to\LCp\antiproton\pip\pim_{\mathrm{non-\SigmaC}}$           & $\left( 39.2 \pm 2.2\comment{_{\mathrm{stat}}} \pm 1.9\comment{_{\mathrm{sys}}} \right)$    & $\left(79  \pm 4\comment{_{\mathrm{stat}}} \pm 4\comment{_{\mathrm{sys}}}  \pm 20\comment{_{\LCp}} \right)$\\[10 pt]
$\Bzb\to\LCp\antiproton\pip\pim_{\mathrm{total}}$           & $\left(61.3  \pm 2.4\comment{_{\mathrm{stat}}} \pm 3.7\comment{_{\mathrm{sys}}} \right)$    & $\left(123  \pm 5\comment{_{\mathrm{stat}}} \pm 7\comment{_{\mathrm{sys}}}  \pm 32\comment{_{\LCp}} \right)$\\
\hline\hline
\end{tabular}
        \label{tab:BranchingFractionsInclSystematics_1}
\end{table*}
\section{Summary}\label{txt:Summary_1}
We observe the decay $\Bzb\to\LCp\antiproton\pip\pim$, study the intermediate decays via $\SigmaC(2455)^{++}$, $\SigmaC(2520)^{++}$, $\SigmaC(2455)^{0}$, and $\SigmaC(2520)^{0}$ resonances, and measure their branching fractions (Table \ref{tab:BranchingFractionsInclSystematics_1}). Yields for events decaying to $\Bzb\to\LCp\antiproton\pip\pim$ without intermediate $\SigmaC(2455,2520)^{++,0}$ resonances are obtained from one-dimensional fits in \minvn, taking information from fits to $\minvn\,{:}\,\mpppm$ into account. For all decay modes, we show the \sPlot distributions of the signal, and we observe significant differences between decays into  $\SigmaCplpl\antiproton\pim$ and $\SigmaCz\antiproton\pip$ final states.

\section{Acknowledgment}
We are grateful for the 
extraordinary contributions of our \pep2\ colleagues in
achieving the excellent luminosity and machine conditions
that have made this work possible.
The success of this project also relies critically on the 
expertise and dedication of the computing organizations that 
support \babar.
The collaborating institutions wish to thank 
SLAC for its support and the kind hospitality extended to them. 
This work is supported by the
US Department of Energy
and National Science Foundation, the
Natural Sciences and Engineering Research Council (Canada),
the Commissariat \`a l'Energie Atomique and
Institut National de Physique Nucl\'eaire et de Physique des Particules
(France), the
Bundesministerium f\"ur Bildung und Forschung and
Deutsche Forschungsgemeinschaft
(Germany), the
Istituto Nazionale di Fisica Nucleare (Italy),
the Foundation for Fundamental Research on Matter (The Netherlands),
the Research Council of Norway, the
Ministry of Education and Science of the Russian Federation, 
Ministerio de Econom\'{\i}a y Competitividad (Spain), and the
Science and Technology Facilities Council (United Kingdom).
Individuals have received support from 
the Marie-Curie IEF program (European Union) and the A. P. Sloan Foundation (USA).

\end{document}